\documentclass[A4Paper,usenatbib,useAMS]{mnras}
\usepackage{natbib}
\usepackage{multirow}
\usepackage{rotating}
\usepackage{pdflscape}
\usepackage{hyperref}
\usepackage{supertabular}
\usepackage[table]{xcolor}

\title[Constraining the assembly time of stellar haloes]{Constraining the assembly time of the stellar haloes of nearby Milky Way-mass galaxies through AGB populations}

\author[Harmsen et al.]{Benjamin Harmsen$^{1}$\thanks{Contact e-mail:\href{mailto:benharms@umich.edu}{benharms@umich.edu}},
						Eric F.\ Bell$^{1}$ \thanks{Contact e-mail:\href{mailto:ericbell@umich.edu}{ericbell@umich.edu}},
						Richard D'Souza$^{2}$, Antonela Monachesi$^{3,4}$,
						\newauthor
						Roelof S.\ de Jong$^{5}$,
						Adam Smercina$^{6}$, In Sung Jang$^{7}$ and Benne W.\ Holwerda$^{8}$
\\
$^{1}$University of Michigan, Department of Astronomy, 311 West Hall, 1085 South University Ave., Ann Arbor, MI 48109-1107\\
$^{2}$Vatican Observatory, 00120 Vatican City State\\
$^{3}$Instituto Multidisciplinario de Investigaci\'on y Postgrado, Universidad de La Serena, Ra\'ul Bitr\'an 1305, La Serena, Chile\\
$^{4}$Departamento de Astronom\'ia, Universidad de La Serena, Av.\ Juan Cisternas 1200 Norte, La Serena, Chile\\
$^{5}$Leibniz-Institut f\"{u}r Astrophysik Potsdam (AIP), An der Sternwarte 16, 14482 Potsdam, Germany\\
$^{6}$Department of Astronomy, University of Washington, Box 351580, Seattle, WA 98195-1580, USA\\
$^{7}$Dept.\ of Astronomy \& Astrophysics, University of Chicago, 5640 S.\ Ellis Avenue, Chicago, IL 60637, USA \\
$^{8}$Department of Physics and Astronomy, 102 Natural Science Building, University of Louisville, Louisville KY 40292, USA\\
}

\hypersetup{draft}
\begin{document}
\label{firstpage}
\pagerange{\pageref{firstpage}--\pageref{lastpage}}
\maketitle

\begin{abstract}

The star formation histories (SFHs) of galactic stellar haloes offer crucial insights into the merger history of the galaxy and the effects of those mergers on their hosts. Such measurements have revealed that while the Milky Way's most important merger was 8-10\,Gyr ago, M31's largest merger was more recent, within the last few Gyr. Unfortunately, the required halo SFH measurements are extremely observationally expensive outside of the Local Group. Here we use asymptotic giant branch (AGB) stars brighter than the tip of the red giant branch (RGB) to constrain stellar halo SFHs. Both stellar population models and archival datasets show that the AGB/RGB ratio constrains the time before which 90\% of the stars formed, $t_{90}$. We find AGB stars in the haloes of three highly-inclined roughly Milky Way-mass galaxies with resolved star measurements from the {\it Hubble Space Telescope}; this population is most prominent in the stellar haloes of NGC 253 and NGC 891, suggesting that their stellar haloes contain stars born at relatively late times, with inferred $t_{90}\sim 6\pm1.5$\,Gyr. This ratio also varies from region to region, tending towards higher values along the major axis and in tidal streams or shells. By combining our measurements with previous constraints, we find a tentative anticorrelation between halo age and stellar halo mass, a trend that exists in models of galaxy formation but has never been elucidated before, i.e, the largest stellar haloes of Milky-Way mass galaxies were assembled more recently.

\end{abstract}


\section{Introduction}

The merging and accretion of galaxies is a central feature of galaxy evolution. Amongst other effects, mergers and accretions are expected to thicken or destroy dynamically-fragile discs \citep[e.g.,][]{Toth1992,Quinn1993,Stewart2008,Hopkins2009}, trigger star formation \citep[e.g.,][]{Barton2003,Chown2019,Moreno2021}, and deliver satellites \citep[e.g.,][]{Deason2015,Patel2020,Dsouza2021}.
The effects of {\it ongoing} mergers can be explored by studying how galaxy properties are different between morphologically-disturbed galaxies, or galaxies in close pairs, and their undisturbed peers \citep[e.g.,][]{Robaina2009,Pipino2014,Violino2018}. Yet, we are also very interested in the {\it long-term} effects of galaxy mergers on Gyr timescales: for example, disc settling and re-growth after mergers \citep[e.g.,][]{Robertson2006,Hopkins2009disk}, or the Gyr-long evolution of satellites after a merger \citep{Deason2015,Dsouza2021,Smercina22}. In order to study these questions, observational measures of the past merger partner(s) of a galaxy --- stellar mass, the merger time, and other quantities --- are needed. Such measures may be accessible by studying stellar haloes --- the accumulated debris from satellite merging and accretion \citep{Bullock01,BJ05}. The stars in a halo are dominated by the contributions of the largest (one or few) merger partners \citep{Cooper10,Deason16,DSouza18a,Monachesi19}, offering a potential probe of past merger mass and time. Unfortunately, these haloes are incredibly diffuse, and even with sensitive integrated light or resolved-star datasets \citep[e.g.,][]{Merritt16,Monachesi16a,Harmsen17,Smercina20}, it is challenging to constrain the merger mass, and the deep resolved-star measurements that are usually used to measure halo age are available only in the Local Group \citep{Gallart2019,Brown06}. In this work, we make progress towards age-dating stellar haloes by quantifying the prominence of bright AGB stars in stellar haloes of three nearby galaxies, calibrating these measurements as crude measures of star formation history, and comparing these measures of halo age to predictions from hydrodynamical models of galaxy formation for the first time. 

The potential scientific pay-offs of such measurements are made clear by our dramatic increase in understanding of the impacts of mergers experienced by the Milky Way and M31, enabled by deep measurements of their stellar haloes. The MW's stellar halo is recognized to be dominated by the debris from a $M_* \sim 10^9 M_{\odot}$ merger that happened around $\sim 8-10$ Gyr ago \citep{Helmi2018,Belokurov2018}. This early and rather low-mass merger coincides in time with the disruption of the MW's early protodisc \citep{Belokurov2020} the formation of the thick disc \citep{Gallart2019}, and appears to have delivered globular clusters \citep{Kruijssen2019}. Such insights are also available for more recent accretions that are separable from the rest of the halo: for example, the Sagittarius dwarf spheroidal galaxy dramatically slowed its star formation as it interacts with the Milky Way \citep[e.g.,][]{Alfaro93,Ruiz-Lara2020}. In contrast, M31's massive, metal-rich stellar halo is dominated by the debris from a much larger merger with a $M_* \sim 1.5 \times 10^{10} M_{\odot}$ {galaxy} $\sim 2$ Gyr ago \citep{DSouza18b,Hammer2018}. This merger is thought to be responsible for M31's disc-wide burst of star formation $\sim 2$ Gyr ago \citep{Williams2015,Williams2017} but did not lead to substantial bulge-building \citep{DSouza18b}, failed to destroy M31's stellar disc but did heat and thicken it \citep{Dorman2015,Hammer2018}, and delivered a substantial fraction of M31's satellites \citep{Weisz2019,Dsouza2021,Savino2022}. The rich learning enabled by knowledge of the MW's and M31's merger and interaction histories, and the hints at the similarities and differences between the effects of the mergers on the host galaxies and their satellites (with a sample size of only two!) clearly motivate the use of such techniques for larger samples of galaxies, reaching into the Local Volume to distances of several Mpc or more, that sample a more representative range of merger histories. 

So what conditions must be met to apply this kind of technique to other galaxies? 

First, it must be possible to either disentangle or ignore(!) the contributions of the numerous disrupted satellite galaxies to a galaxy's stellar halo. Fortunately, this condition is largely met. For systems with prominent streams (e.g., M31; \citealt{Ibata2001}, M83; \citealt{MalinHadley97}) targeted observations of the streams {will constrain the mass, metallicity and star formation history} of that particular interaction (e.g., \citealt{Brown06}). {Yet, we generally want to know not only about recent events but also the most massive merger, which may have been earlier.} The debris from more ancient mergers and accretions is phase mixed and cannot be disentangled spatially. Fortunately, owing to the steepness of the stellar mass--dark matter halo mass correlation, the largest single merger generally contributes most of the stars to a given halo \citep{Cooper10,Deason16,DSouza18a,Monachesi19}. This manifests itself observationally in a relatively tight relationship between the metallicity and stellar mass of a stellar halo --- more massive haloes have higher metallicities, owing to most of their stars having come from a more massive, more metal-rich progenitor (e.g., \citealt{Harmsen17}, \citealt{DSouza18a}, \citealt{Monachesi19}). Consequently, the star formation history of the well-mixed part of the halo will constrain the star formation history of the galaxy's largest merger. 

Second, there should be a clear observational star formation history signature corresponding to the accretion of a galaxy. Overall, it is clear that the ultimate effect of merging and accretion is a shut-down of star formation in the accreted satellite, through a combination of gas loss by interaction with the main galaxy's circumgalactic medium (ram pressure stripping acting in concert with other processes; e.g., \citealt{Mayer2006}, \citealt{Slater2014_sat}) and the final tidal disruption of the secondary, which by necessity reduces gas densities to the point where the secondary no longer forms stars. The balance of these processes largely depends on the mass of the secondary: at low masses ($< 10^8 M_{\odot}$), the statistics of star formation in nearby satellite galaxies suggests that ram pressure of the main galaxy's circumgalactic medium is important to largely shut down star formation around the time of infall, while at higher masses (similar to the Magellanic Clouds) the gas stays in the satellite for longer \citep[e.g.,][]{Slater2014_sat,Fillingham2015,Fillingham2016}. Such behaviour is clearly seen in hydrodynamical models of galaxy formation (\citealt{Samuel2022}; see also \S \ref{subsec:TNG}). While there are uncertainties about the correspondence between star formation shut-down time and important orbital events (e.g., infall, first pericenter, final disruption; \citealt{Weisz2015}) in detail, it is nonetheless clear that the time at which star formation shuts off in a halo will give a measure of merger or accretion time. 

Third, one must have access to measures of star formation shut-off with sufficient sensitivity to meaningfully constrain past mergers and interactions. While in principle spectral signatures may be helpful towards this goal \citep[e.g.,][]{Johnson2021}, haloes and streams have sufficiently low surface brightness as to make this approach unfeasible. Instead, resolved star approaches are necessary. High-fidelity SFH measurements for ancient ($>4-6$\,Gyr) populations requires photometric depths reaching below the main sequence turn-off of old stellar populations --- a condition only met well within the Local Group (e.g., \citealt{Brown06}, \citealt{Gallart2019}). Shallower photometry reaching below the Horizontal Branch/Red Clump can still measure SFH, and when star formation shuts off, with reasonably high fidelity for intermediate ages, but errors grow very large for populations with ages $>4-6$\,Gyr \citep{Weisz2011}. 

Outside the Local Group, it is impossible to reach the old star main sequence turn-off. Horizontal Branch depth data exists for only two stellar haloes of galaxies with masses similar to the Milky Way --- deep halo pointings in M81 \citep{Durrell10} and NGC 5128/Centaurus A \citep{Rejkuba2005} for which the morphology of the RGB, the Red Clump, the RGB bump and the AGB bump can jointly constrain the SFH, and in particular the star formation shut-off time. On this basis, the bulk of stars in M81's halo population are found to be $9\pm2$\,Gyr old \citep{Durrell10}. While the average age of NGC 5128's halo is a little younger at $8^{+3}_{-3.5}$\,Gyr \citep{Rejkuba2005}, the CMD morphology demonstrates that substantial star formation continued until around $t_{\rm shut\,off}\sim2$\,Gyr ago \citep{Rejkuba2011}. Yet, resolved star data exists for many more galaxies to shallower limits, probing the upper 1-3 magnitudes of the red giant branch using both HST \citep[e.g.,][]{RadSmith11,Monachesi16a,Harmsen17,Cohen15,Peacock15} and from the ground \citep[e.g.,][]{Greggio14,Crnojevic16,Smercina20}. Some of those studies have noted that some stellar haloes have bright AGB stars --- those brighter than the tip of the red giant branch \citep{Greggio14,Rejkuba2022}. These bright AGB stars are expected to be particularly prominent in systems with a significant mass in intermediate-age stellar populations, and are much less numerous in old populations, offering complementary insight to the metrics available for deeper magnitude limits. {While there are still very important uncertainties in this phase of AGB star evolution (e.g., \citealt{Marigo2017,Pastorelli2019}, see also Appendix \ref{ap:agb_unc}), this nonetheless} opens the possibility of using bright AGB stars as a tracer of star formation history, and star formation shut-off time in particular, for a broader sample. Indeed, in the case of Cen A, these AGB stars -- and particularly the ratio of AGB to RGB stars --- were used to confirm a $t_{\rm shut\,off}\sim2-3$\,Gyr for Cen A's stellar halo \citep{Rejkuba2022}.  

In this paper, we further explore the utility of bright AGB stars --- those brighter than the tip of the red giant branch --- as crude tracers of the star formation shut down time in stellar halo datasets. As an initial step in developing this technique for halo study, we focus our efforts on quantifying the bright AGB/RGB ratio (and how it varies from place to place) in three nearby MW-mass galaxies (M81, NGC 253 and NGC 891) with the required quality of data from the GHOSTS (Galaxy Haloes, Outer discs, Substructure, Thick discs, and Star clusters) survey \citep{RadSmith11} carried out with {\it HST}. By comparing this ratio with stellar population models and inferred SFHs for observed dwarf galaxies, we are able to calibrate the AGB/RGB ratio as an approximate measure of when star formation shuts off in a halo (or a halo feature), show that halo substructures are younger than the well-mixed haloes, and reveal an anti-correlation between stellar halo `age' and stellar halo mass that has not been seen before but is a prominent feature of simulated stellar haloes in hydrodynamical simulations of galaxy formation.

\section{Observations, Data Reduction and Photometry}

\subsection{Observations and selection of galaxies for study}

Our observations were taken from the GHOSTS {(Galaxy Haloes, Outer discs, Substructure, Thick discs, and Star clusters)} survey \citep{RadSmith11}. The survey resolves individual stars in 18 Local Volume disc galaxies (with a range of masses and inclinations) using the HST Advanced Camera for Surveys (ACS) and Wide Field Camera 3 (WFC3) using the F606W and F814W filters. We focus on a subset of these galaxies: those that are highly inclined, to ensure that at least the minor axis fields are dominated by stellar halo material; and the subset with stellar masses comparable to the Milky Way (as presented previously by \citealt{Monachesi16a} and \citealt{Harmsen17}). From this subset, we narrow our consideration further to three galaxies where we can robustly measure (or place constraints on) the AGB and RGB populations: M81, NGC 253 and NGC 891. Of the sample studied by \citet{Monachesi16a} and \citet{Harmsen17}, NGC 4945 is at sufficiently low galactic latitude that its AGB population is undetectable against the much more numerous Milky Way foreground stellar population. NGC 4565 has a widespread young population (visible as blue stars in the CMDs of fields 5 and 6 in Fig.\ 4 of \citealt{Monachesi16a}); metal-poor bright AGB stars are in a region of the CMD close to the red supergiant/core helium burning stars that are prominent in stellar populations with ages $<200$\,Myr, and so we choose for simplicity not to analyse its AGB population. NGC 7814 is sufficiently distant that while its AGB population is well-measured, its RGB population is incomplete, particularly for metal-richer stars; again, we choose not to analyse it further here. 

\begin{figure*}\centering
\includegraphics[width=170mm]{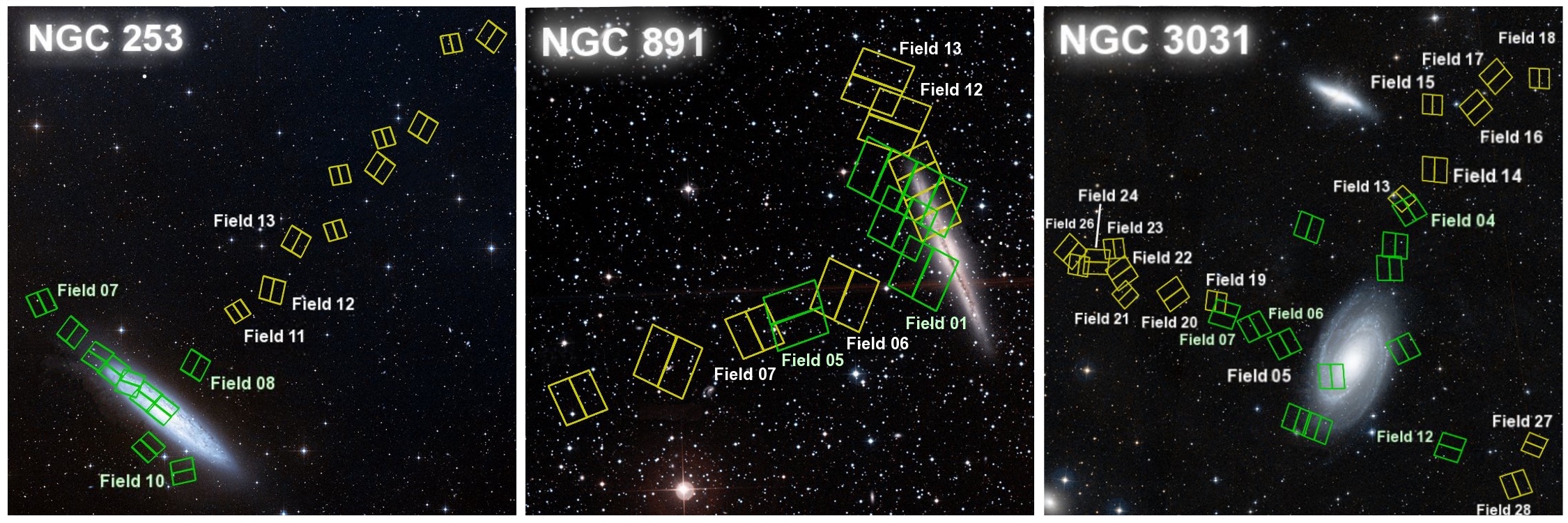}
\caption{Location of the GHOSTS HST ACS/WFC and WFC3/UVIS fields overlaid on images from the Digitized Sky Survey. Fields introduced in \citet{RadSmith11} are coloured green, while fields introduced in \citet{Monachesi16a} are coloured yellow. {We do not use all fields in this work; the fields used here are labeled with field numbers and are enumerated in Table \protect\ref{Composite CMDs Table}}. The images are oriented such that north is up and east is to the right.}
\label{fields}
\end{figure*}

This leaves a sample of three galaxies for this initial study: M81, NGC 253 and NGC 891. We show the layout of the HST survey fields in Fig.\ \ref{fields}. {Note that some of the fields identified in Fig.\ \ref{fields} are not used in this study; the fields that we analyse are labeled in Fig.\ \ref{fields} and enumerated in Table \ref{Composite CMDs Table}}. We adopt distances and corresponding distance moduli from \citet{RadSmith11}: $D=3.5$\,Mpc ($m-M=27.70)$ for NGC 253, $D=9.1$\,Mpc ($m-M=29.80)$ for NGC 891, and $D=3.6$\,Mpc ($m-M=27.79)$ for M81.

\subsection{Data Reduction and Photometry}
\label{Data Reduction and Photometry}
Here we will provide a summary of the data reduction process outlined in \cite{Monachesi16a} and \cite{RadSmith11}. We download or generate $*$\verb+_flc+ from the Hubble Data Archive MAST \footnote{\url{http://archive.stsci.edu}}. The images are bias-subtracted, flat fielded, corrected for charge transfer efficiency \citep[CTE;][]{Anderson_bedin10}, and as many cosmic rays are removed as is possible. The individual $*$\verb+_flc+ files were combined and corrected for geometric distortion using Astrodrizzle \citep{Gonzaga12}. 

For star detection and photometry, we used DOLPHOT, which is an updated version of HSTphot \citep{Dolphin00}. DOLPHOT utilises point-spread function (PSF) fitting of sources in the FLC images to provide accurate PSF photometry for sources. DOLPHOT occasionally will detect spurious point sources in image artefacts or the outer parts of bright stars or galaxies. We use SE\verb+XTRACTOR+ \citep{Bertin_arnouts96} to detect bright, extended sources and construct a mask that encloses the light from these bright sources. This generated mask gives the unusable area of the ACS or WFC3 field, consisting of background galaxies, foreground stars, globular clusters, or bad pixels. Point source detections were removed if they were within a certain distance of the masked area, 25 pixels for ACS and 5 pixels for WFC3. Cosmic rays were also masked out in a process detailed in \cite{Monachesi16a}. Magnitudes are given in this work in the Vega magnitude system. 

At the brightnesses of interest for this work ($m_{\rm F814W} > 22$), compact background galaxies are an important contaminant, even at HST resolution. In order to reject these compact background galaxies, we use a combination of criteria on the signal to noise and shape of the detected sources to reject likely contaminants (we refer to this as `culling'; a cursory overview will be provided here, for more details, see \citealt{RadSmith11} and \citealt{Monachesi16a}). 
{For ACS/WFC culls, we applied these criteria:\break

{\small
\begin{equation}
    \mathrm{-0.06<SHARPNESS_{F606W}+SHARPNESS_{F814W}<1.30 }
\end{equation}
\begin{equation}
    \mathrm{CROWDING _{F606W} + CROWDING_{F814W}<0.16}
\end{equation}
\begin{equation}
    \mathrm{S/N _{F606W}>5.0,\hspace{0.5cm}S/N_{F814W}>5.0}
\end{equation}

\noindent
The culls for WFC3/UVIS were the following:
\vspace{5mm}
\begin{equation}
    \mathrm{-0.06<SHARPNESS_{F606W}+SHARPNESS_{F814W}<1.50 }
\end{equation}
\begin{equation}
    \mathrm{CROWDING _{F606W} + CROWDING_{F814W}<0.20}
\end{equation}
\begin{equation}
    \mathrm{S/N _{F606W}>5.1,\hspace{0.5cm}S/N_{F814W}>3.2}
\end{equation}}
}
When this process is carried out in empty fields, 95\% of the contaminants are rejected and only a modest number of sources is left behind --- faint halo stars in the Milky Way, and a few residual background sources that are unresolved at HST resolution and our S/N. In order to quantify the likely influence of these sources, we follow \citet{InSung2020M101} and analyse four `empty' fields for each of ACS and WFC3 in the same way as this dataset to quantify the background and its variation from place to place on the sky. 

We determine our photometric uncertainty/completeness by running artificial star tests. A total of 2,000,000 fake stars with a reasonable variation in colour and magnitude are analysed for each field, using the same DOLPHOT culls as the HST data \citep{RadSmith11,Monachesi16a}. The ratio between stars that pass the culls and the total number of injected stars gives us our completeness ratio. Completeness varies within fields, mostly as a function of brightness, and between fields. 

\section{Description of data analysis}
\label{Data Analysis}

In this section, we describe the most important features of our data analysis: the motivation and choice of our AGB and RGB star selection regions on the colour--magnitude diagram, the areas over which we aggregate to give as well-measured as possible AGB/RGB ratios, and a discussion of background subtraction and our sources of uncertainty. 

\subsection{Choice of AGB and RGB colour-magnitude diagram selection regions}

\begin{figure*}\centering
\includegraphics[width=190mm]{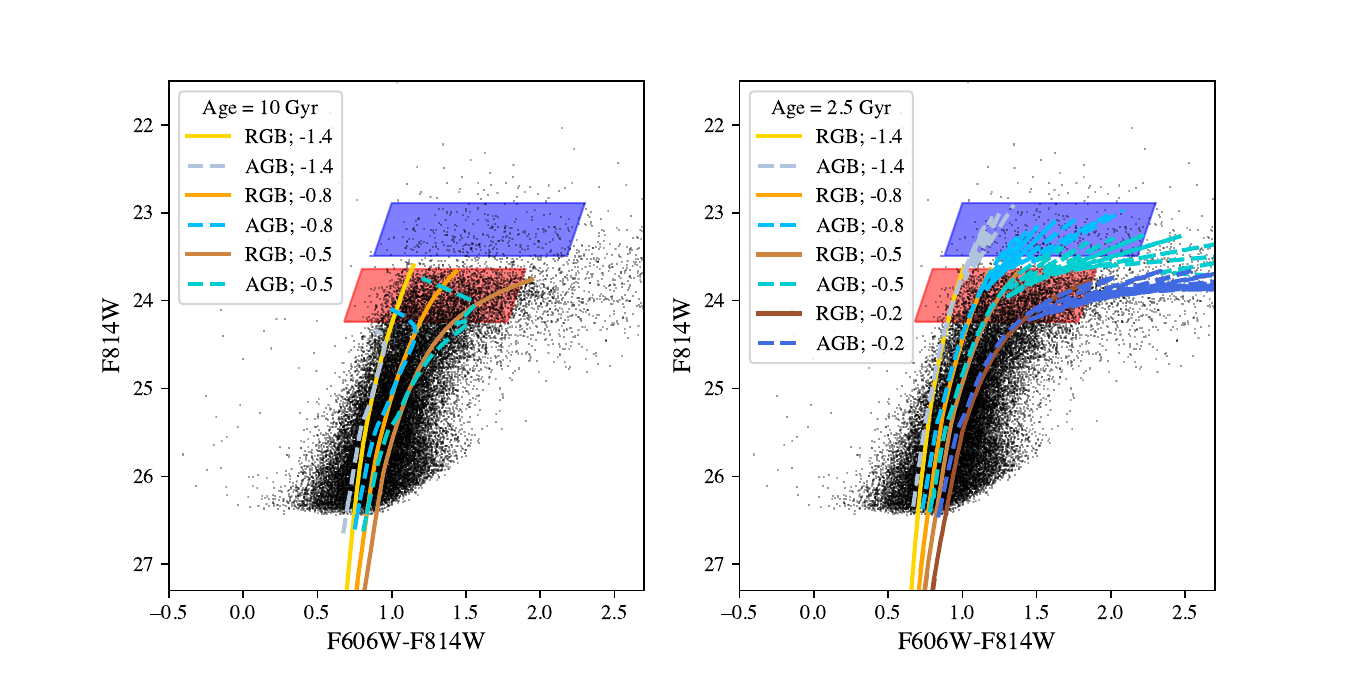}
\caption{Colour--magnitude diagram for GHOSTS stars in NGC 253's inner stellar halo, with {Padova CMD v3.2} isochrones for two different ages (left: 10Gyr; right: 2.5Gyr) for a range of different metallicities ([M/H]$=-1.4$ to [M/H]$=-0.5$ in the left panel, and [M/H]$=-0.2$ in the right panel. RGB isochrones are in shades of yellow and brown, and AGB isochrones are shown as dashed lines in shades of blue. The selection regions that we adopt in this work are highlighted in red (RGB) and blue (AGB). }
\label{AGBRGB_iso}
\end{figure*}

We describe first our selection criteria for AGB and RGB stars. We start by noting that we have confirmed that the conclusions of our work are not affected qualitatively by the detailed choice of AGB and RGB selection box --- the most important thing to bear in mind for analyses of this type is to analyse the stellar halo and calibration data (isochrones, reference galaxies) using the same selections. With this consideration in mind, we now describe our selections and the rationale for the choices that we have made. 

We illustrate our selections in Fig.\ \ref{AGBRGB_iso}, where we show a sample of GHOSTS stars in the halo of NGC 253, along with a range of isochrones at two reference ages (10 Gyr on the left, and 2.5 Gyr on the right). The model stellar populations use Padova isochrones, {CMD version 3.2} \footnote{\url{http://stev.oapd.inaf.it/cmd}} using the PARSEC evolutionary tracks version 1.2S for all phases of stellar evolution except for the AGB phase \citep{Bressan2012,Chen2014,Chen2015,Tang2014}, and the COLIBRI AGB evolutionary tracks including the thermally-pulsing AGB phase following \citet{Marigo2017}. {The thermally-pulsing AGB phase is subject to very important modeling uncertainties; we discuss this further in Appendix \ref{ap:agb_unc}}. This visualization makes the motivating idea of this analysis clear --- old populations (left) are expected to have very few bright AGB stars, whereas younger populations (e.g., 2.5 Gyr on the right) have {thermally-pulsing} AGB stars that exceed TRGB stars in brightness during some of their thermally pulsing phase. Some haloes (NGC 253's in this case) have a substantial population of AGB stars brighter than the TRGB, implying an extended star formation history for material that ends up in its stellar halo. 

In order to be able to inter-compare AGB/RGB ratios across our sample, and calibrate to measures of SFH using theoretical models and observational data, we define a single selection box for all galaxies that is referenced to the brightness of the TRGB. Our selections (the upper box is shaded blue for AGB, the lower box is shaded red for RGB) cover a wide range of metallicities (up to [M/H]$=-0.5$ at old ages, and even somewhat higher metallicities for younger ages), particularly for the RGB phase. The RGB selection uses a box that extends 0.6 magnitudes below the TRGB; this limit avoids significant incompleteness even for our most distant galaxy, NGC 891. The AGB selection box also has a depth of 0.6 mag, with the brightest edge located 0.75 mag above the TRGB. The colour ranges are chosen with a few competing considerations in mind. At the blue edge, the limit excludes bluer Main Sequence and {blue Helium-Burning stars, while tracking trends expected from the RGB and AGB isochrones. We note that some red Helium-Burning stars could be included in the bluest colour edge --- while this is not a concern for our galaxies and regions of consideration, for systems with a rich red Helium-Burning star population, one would want to adopt a slightly redder cut at the expense of losing a few RGB and AGB stars}. The width is chosen to encompass the relevant range of metallicities for stellar haloes (making AGB/RGB ratio insensitive to metallicity variations between haloes; \S \ref{subsec:ratio_mod}) while steering clear of the 70\% completeness limits for very red stars (especially for our most distant galaxy, NGC 891) and incurring excessive contamination from red foreground low-mass Galactic stellar halo stars (especially visible for our calibration sample of galaxies, e.g., \S \ref{subsec:ratio_obs}). 
The corners for the RGB selection box (relative to the apparent magnitude of the TRGB) are located at $(0.8, 0)$ $(1.9, 0)$ $(0.68, 0.6)$ $(1.78,0.6)$, and for the AGB selection box $(1.0, -0.75)$ $(2.2, -0.75)$ $(0.88, -0.15)$ $(2.08, -0.15)$ (see Fig.\ \ref{AGBRGB_iso}). Given $M_{\rm F814W,TRGB} = -4.06$, we assume that the TRGBs are at $m_{\rm F814W} = 23.64$, 25.74 and 23.73 for NGC 253, 891 and 3031 respectively (following \citealt{RadSmith11}). 

\subsection{Choice of regions for consideration} \label{aggregation}

\begin{figure*}\centering
\includegraphics[width=190mm]{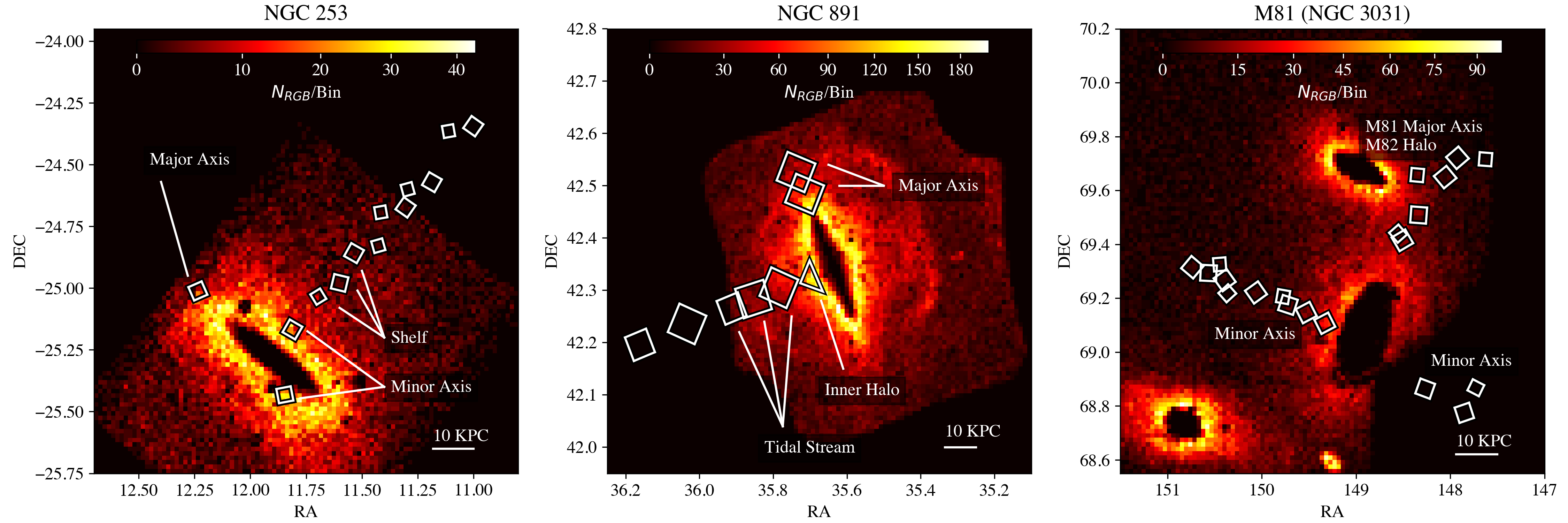}
\caption{RGB density map with HST fields overlaid, showing the substructure present in the galaxy haloes and where it appears in our chosen HST fields. Panoramic data for NGC 891 and M81 are from Subaru, while the data for NGC 253 is from the VISTA telescope. For NGC 891, Field 1 has been adjusted to show the area used in our investigation, as the stars contained in the innermost region of Field 1 were not used. For NGC 253 and NGC 891, the fields that fall within the shelf and tidal stream are labelled. The bin size for each galaxy is approximately 1.5\,kpc$^2$ for NGC 253, 2\,kpc$^2$ for NGC 891, and 1.2\,kpc$^2$ for M81.
\label{substructuremap} }
\end{figure*}

AGB stars, particularly for older populations, are quite rare, requiring a substantial mass in stars to be surveyed to place useful constraints on AGB/RGB ratio. In order to measure AGB/RGB ratio to an accuracy of 0.05 or 0.15\,dex respectively, roughly 100 or 10 AGB stars are needed. For older populations with $\log_{10} N_{AGB}/N_{RGB} \sim -1.4$, this requires $10^8 M_{\odot}$ (or $10^7 M_{\odot}$) respectively, corresponding roughly to $M_V \sim -15$ or $-12.5$. For younger populations with $\log_{10} N_{AGB}/N_{RGB} \sim -1.0$, this would require $2.5\times 10^7 M_{\odot}$ (or $2.5\times 10^6 M_{\odot}$), or $M_V \sim -14$ or $-11.5$. In most cases, we needed to combine multiple GHOSTS fields to reach the number of AGB stars required to determine AGB/RGB ratios to useful accuracy. {Because stellar halos have substructure \cite[e.g.,][]{BJ05,Monachesi19}, such aggregation will by necessity involve some averaging away of spatial variations in star formation history and AGB/RGB ratio.  }

{Therefore,} bearing in mind the possibility of spatial variations in AGB/RGB ratio, we experimented with different ways to combine fields.  In the end, we found it most useful to classify the fields in our three galaxies into 3 categories: inner halo, substructure, and major axis fields. The inner halo are those fields close to the minor axis, lying more than 7\,kpc from the centre of the galaxy to avoid important contributions from {\it in situ} disc stars, with no obvious substructure. For NGC 253 and NGC 891, these are the innermost minor axis halo fields. NGC 891 is distant enough that each HST field covers 12.5\,kpc corner-to-corner; we consider only the outer part of Field 1,  defined as being at least 7 kpc from the galaxy centre, for the inner halo to steer well clear of the disc plane. For M81, all minor axis fields were included. 
We consider separately those fields that have stellar halo overdensities that were already known from the panoramic or HST star count studies. In the case of NGC 253, \citet{Greggio14} revealed an overdensity at $\sim 30$\,kpc distance along the northwestern minor axis (see, e.g., their Fig. 16), also reproduced as a flattening (or shelf) in the minor axis star counts power-law profile from \citet{Harmsen17}. We indicate those shelf fields in the left-hand panel of Fig.\ \ref{substructuremap}. We note also that the inner stellar halo fields include material associated with NGC 253's inner halo shell \citep{MalinHadley97,Bailin11,Greggio14}, which overlaps particularly with Field 10; we saw no evidence of spatial variations in AGB/RGB ratio within the inner stellar halo, and so we consider Fields 8 and 10 together in what follows as NGC 253's inner stellar halo. 
NGC 891 hosts clear tidal streams in its halo \citep{Mouhcine10}, visible again as a deviation from a power-law profile in \citet{Harmsen17}. We identify the fields with important contributions from this tidal stream material in the centre panels of Fig.\ \ref{substructuremap}. In the case of M81, the panoramic map of \citet{Smercina20} combined with the major axis star counts from \citet{Harmsen17} make it clear that all of the major axis fields that we consider have very important contributions from the outer parts of M82's halo and tidal debris field{; accordingly, we refer to M81's major axis field as `M81 major axis$+$M82 halo'.} Finally, {NGC 253's and NGC 891's} major axis fields are considered separately, in part owing to a recognition that it is much less clear how important {\it in situ} or migrated disc stars might be to those fields (see, e.g., \citealt{Pillepich15}, \citealt{Monachesi16b}).  We give the classification of the fields into these three categories in Table \ref{Composite CMDs Table} and a visual guide in Fig.\ \ref{substructuremap}.

\begin{table}\centering
    \begin{tabular}{l l l}
    
    Galaxy      &Region       &Fields\\
    \hline
    NGC 253     &Minor Axis (Inner Halo) & 8, 10\\
                &Major Axis &  7\\
                &Shelf & 11, 12, 13\\
               \hline
    NGC 891     &Minor Axis (Inner Halo) & 1 (outer portion)\\
                &Tidal Stream& 5, 6, 7\\
                &Major Axis & 12, 13\\
               \hline
    M81 (NGC 3031)     &Minor Axis &5, 6, 7, 12, 19, 20, 21\\
                &           &22, 23, 24, 26, 27, 28\\
                &Major Axis$+$M82 Halo  &4, 13, 14, 15, 16, 17, 18\\
                                 
    \end{tabular}
    \caption{Table of the fields used and their classification.}
    \label{Composite CMDs Table}
\end{table}

\subsection{Determination of AGB/RGB ratios, background subtraction, the impact of blending, and uncertainties}

We normalize the AGB and RGB star counts to a surface density by dividing by the area of each HST field, determined by summing their un-masked pixel areas; this allows us to combine measurements from both cameras and correctly account for the differing fraction of areas masked in each exposure. We account for completeness using the artificial star tests. For each un-masked star that has passed our culls, we determine the ratio of injected/recovered artificial stars that are nearby in both location on the field and in colour magnitude space: for stars in our selection regions, more than 70\% of the injected stars are recovered in all cases. This gives an estimated `intrinsic' number of stars in our CMD selection box in that region of the field that correspond to each observed star, allowing us to calculate the AST-corrected star counts and surface densities for each field. Uncertainties were estimated by bootstrapping the stars in each CMD selection region for each aggregated area. 

The AGB and RGB stellar densities for each region have contributions from a combination of foreground Galactic stars and unresolved background galaxies. We follow \citet{InSung2020M101} in using a set of `empty' high Galactic latitude fields with ACS and WFC3/UVIS coverage in the F606W and F814W filters as control fields. The fields were chosen to be far away from bright stars, galaxies, tidal streams, star clusters, or galaxy clusters. We analyse these fields using the same pipeline as the other GHOSTS fields, applying our masking around spurious sources and culls in exactly the same way as we do for our dataset. We use 4 empty fields for each of the ACS and WFC3, and apply our CMD boundaries (which are specific to each of NGC 253, NGC 891 and M81) to the detections resulting in 8 background values for AGB stars and RGB stars separately. We find no systematic difference between the backgrounds derived from ACS and WFC3; accordingly, we average these eight values to estimate the background density for each of the AGB and RGB selections for our three target galaxies. In order to account for possible variation from field to field in the background density, we bootstrap these eight fields to give an estimate of the uncertainty in the background density; indeed, models of Galactic foreground star contamination suggest relatively little variation in the inferred foreground star counts from field-to-field (see Appendix \ref{ap:fgbgcont}). Our results are not especially sensitive to the choice of background subtraction strategy --- background subtraction using the outermost fields of each galaxy clearly modestly oversubtracts (especially the more numerous RGB star population), but gives AGB/RGB values similar to those presented here. 

A final consideration is unrecognised blending of bright RGB stars that might appear as a single, brighter AGB star. Given the probability $P_{RGB}$ of a bright RGB star per unit area (which we choose to be the area of the F814W FWHM, a circle 0$\farcs$09 in diameter), the probability of a blend will be $P_{RGB}^2$. This clearly depends on the density of RGB stars, which is low in the stellar halo fields that we analyse. One can estimate a rough {\it upper limit} on blending by analyzing the two dense NGC 253 inner halo fields (Fig.\ \ref{AGBRGB_iso}); $<1$\% of the AGB stars are expected to be blends, even in these dense and crowded fields. We conclude that blending is unimportant at the stellar densities characteristic of stellar halo fields.

The final AGB/RGB ratios are determined by dividing the 
AST-corrected and background subtracted AGB and RGB star counts for the areas under consideration. As described earlier, owing to the rarity of AGB stars, we generally consider areas that are aggregates of multiple HST fields. Our composite CMDs for each aggregated region are shown in Fig.\ \ref{Composite CMDs}, and our AGB/RGB ratios and estimates are given in Table \ref{AGB RGB Ratios Table}.

\section{Results}
\label{Results}

\begin{figure*}\centering
	\includegraphics[width=170mm]{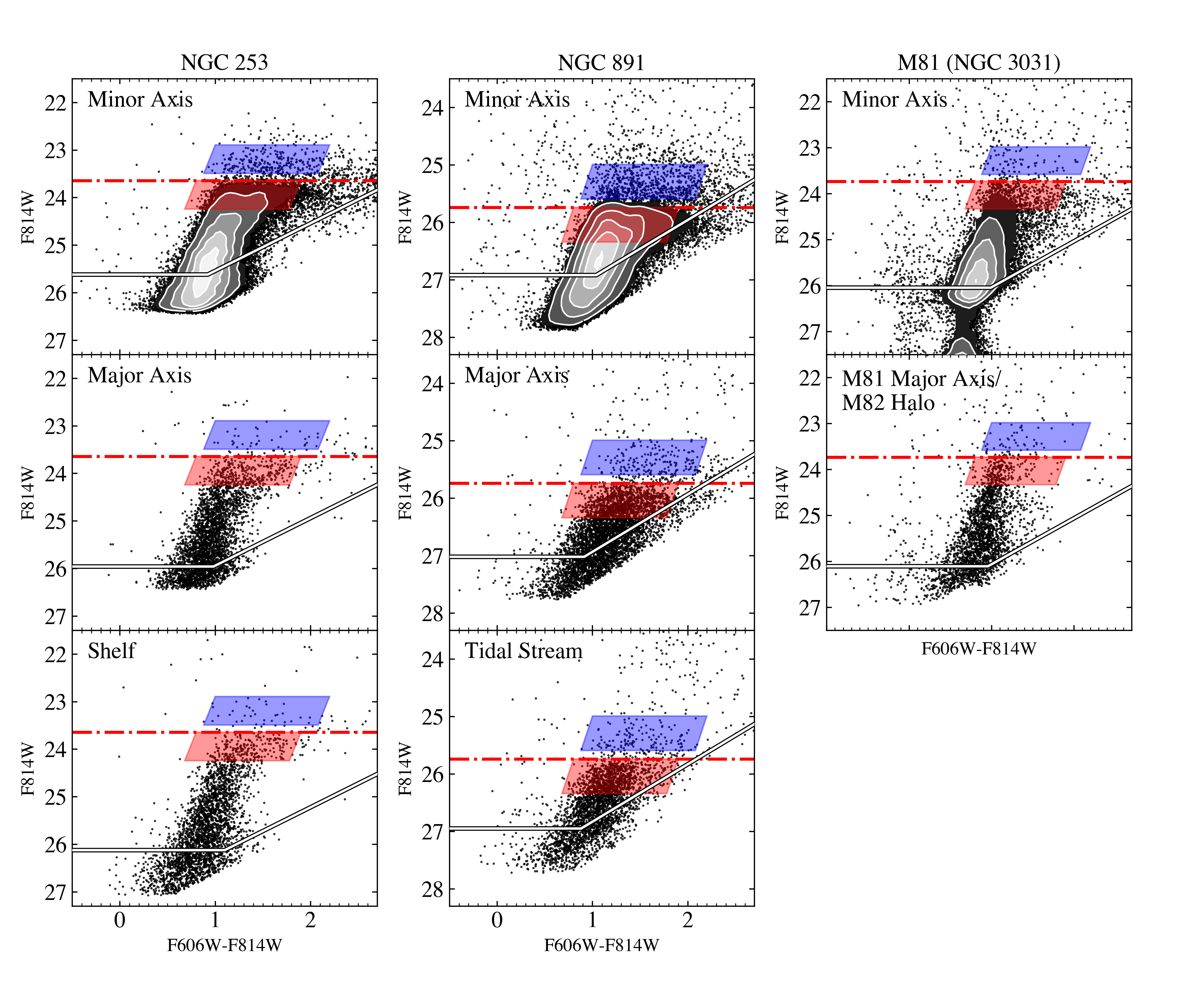}    
	\caption{The colour-magnitude diagrams for NGC 253, NGC 891 and M81, for the `minor axis', `major axis' and substructure regions for each galaxy (Table \protect\ref{Composite CMDs Table}). The stars from all fields in their respective regions are plotted together. The
	red (lower) box outlines the selection cuts for RGB stars, the blue (upper) box
	outlines the cuts for AGB stars. The dotted/dashed line is the TRGB, and the
	white/black line marks the 70\% completeness limits as determined by the artificial star tests. Only
	detections that passed photometric culls outlined in {\protect\cite{RadSmith11}} and
	{\protect\cite{Monachesi16a}} are shown. One of the fields in NGC 3031's halo is substantially deeper than the rest (from \citealt{Durrell10}), thus reaching unusually faint magnitudes, which can be seen in the Minor Axis panel. This effect is demonstrated to a lesser degree in NGC 891's tidal stream and NGC 253's shelf.  
	}
	\label{Composite CMDs}
\end{figure*}

Figure \ref{Composite CMDs} shows the composite CMDs, aggregated by region (to simultaneously maximise the number of relatively rare AGB stars on one hand while being able to probe for spatial variations on the other), for each of our three galaxies. As described earlier, we adopt identical selection regions (relative to each galaxy's TRGB) for the AGB and RGB regions. 

Each galaxy's CMD shows a prominent RGB, tracing the older stellar populations that are dominant in stellar haloes, and as analyzed by \citet{Monachesi16a} and \citet{Harmsen17}. For NGC 253 and NGC 891, a significant number of AGB stars are visible that are brighter than the TRGB; these are especially prominent in the CMD for NGC 253's minor axis. M81 shows a less apparent bright AGB star population. In addition, a prominent young ($<$200\,Myr old) stellar population is detected in M81 bluewards of the RGB and AGB, representing stars formed in M81's group-wide {\sc Hi} debris field \citep{Okamoto15,Okamoto2019}. No such population is prominent for NGC 253 and NGC 891. In addition, at brighter magnitudes (brighter than the AGB), M81 and NGC 891 show a modest contribution from foreground Galactic stars, owing to their somewhat lower galactic latitude. 

\begin{figure*}
\centering
\includegraphics[width=180mm]{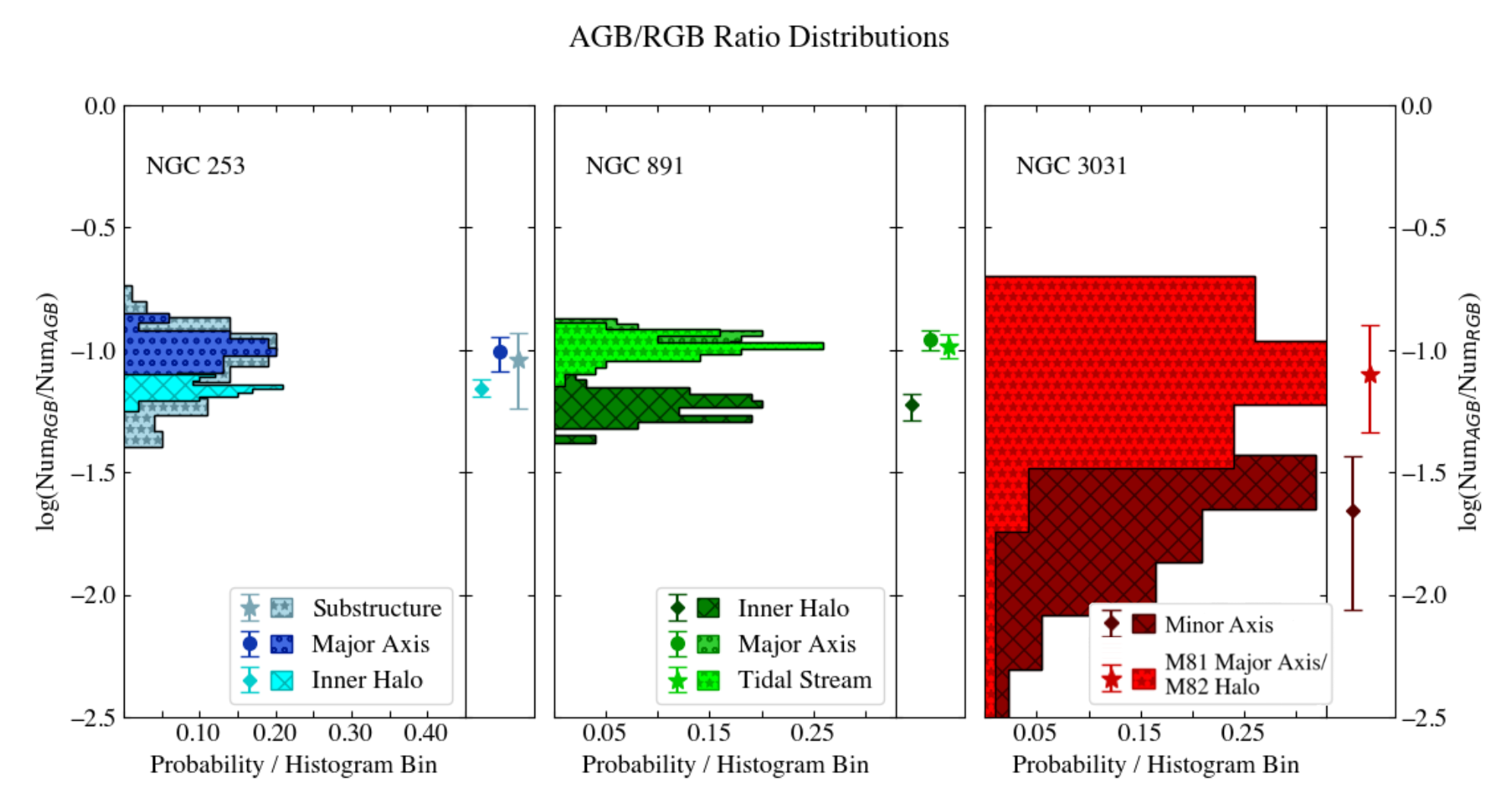}
\caption{
Displayed here are three panels corresponding to the three GHOSTS galaxies in our sample. Each panel consists of histograms showing the distribution of AGB/RGB ratios derived from bootstrapped samples of the data (left section), in addition to the resulting AGB/RGB ratios including $\pm1\sigma$ confidence intervals (right section). The y axes have matching scales to make clear the differences in AGB/RGB ratios between galaxies. 
The x axis corresponds to probability per bin, which is the number of values in a bin divided by the total number of values across all bins. The right section of each panel serves to show the resulting ratio and its error bar and allow for clear comparison between galaxies.}
\label{Ratios Histogram Plot}
\end{figure*}

\begin{table*}
\centering
    \def\arraystretch{1.1}
    \begin{tabular}{|c|ccc|ccc|cc|}
                \hline
                Galaxy      
                &\multicolumn{3}{c|}{log$(N_{AGB}/N_{RGB})$}
                &\multicolumn{3}{c|}{Inferred $t_{90}$ (Gyr)}
                &Stellar Mass&Stellar Halo Mass \\
                \hline
                NGC 253 & Inner Halo & Shelf & Major Axis & Inner Halo & Shelf & Major Axis & &\\
                & $-1.16^{+0.04}_{-0.03}$ 
                & $-1.04^{+0.11}_{-0.20}$ 
                & $-1.01^{+0.08}_{-0.07}$ 

                & $6.2 \pm 1.5$
                & $4.9^{+2.4}_{-2.1}$ 
                & $4.5\pm1.7$ &  $5.5\pm1.4\times10^{10}$ & $4.5^{+0.5}_{-0.3}\times10^9$   \\
                \hline
                NGC 891 & Inner Halo & Tidal Stream & Major Axis & Inner Halo & Tidal Stream & Major Axis & &\\
                & $-1.22^{+0.05}_{-0.07}$ 
                & $-0.99^{+0.05}_{-0.04}$ 
                & $-0.96\pm0.04$ 

                & $6.9\pm1.6$
                & $4.0 \pm 1.5$
                & $4.2 \pm 1.5$
                & $5.3\pm1.3\times10^{10}$ & $2.68^{+0.22}_{-0.16}\times10^9$   \\
                \hline
                M81 & Minor Axis & \multicolumn{2}{c|}{Major Axis$+$M82 Halo} & Minor Axis & \multicolumn{2}{c|}{Major Axis$+$M82 Halo} & & \\
                & $-1.65^{+0.21}_{-0.41}$ 
                & \multicolumn{2}{c|}{$-1.10^{+0.20}_{-0.24}$}  

                & $11.8\pm4.0$
                & \multicolumn{2}{c|}{$5.3\pm2.5$} 
                & $5.6\pm1.4\times10^{10}$ & $1.14^{+0.11}_{-0.07}\times10^9$   \\
                \hline
    \end{tabular}
    \caption{The AGB/RGB ratio for the different regions of our selected galaxies, along with inferred $t_{90}$ values for each of those regions {(see Section \protect\ref{subsec:ratio_obs}, Equation \protect\ref{eq:t90})}. The total stellar mass and estimated total stellar halo masses (in solar masses) of galaxies are also given for the reader's convenience, following \protect\cite{Harmsen17}. 
    %last updated by Eric $2022:10:11$
    }
    \label{AGB RGB Ratios Table}
\end{table*}

We show the distribution of AGB/RGB ratios for each region in Fig.\ \ref{Ratios Histogram Plot}, and their medians and $\pm 1 \sigma$ confidence intervals, as inferred using the bootstrapping analysis described in Section \ref{Data Analysis}. The numerical results are provided in Table \ref{AGB RGB Ratios Table}.
A cursory glance at the results shows variation in the ratio between galaxies and within them, in different regions of the halo. In the following section we will explore some of the general trends we find while also taking a closer look at each galaxy individually. These results will then be interpreted in Section \ref{Interpretation}.

\subsection{NGC 253}

While the AGB/RGB ratio for the minor axis ($-1.16^{+0.04}_{-0.03}$) of the inner stellar halo is very well measured, the modest number of HST fields coupled with the small area covered by each HST field (at the distance of NGC 253, $D=3.5$\,Mpc) limit the accuracy of our other measurements. NGC 253's shelf substructure \citep{Greggio14,Harmsen17} has an AGB/RGB ratio ($-1.04^{+0.11}_{-0.20}$) which appears somewhat higher than its inner stellar halo, although with uncertainties that overlap the inner stellar halo's value. NGC 253's major axis, comprising field 7, shows a very similar ratio of $-1.01^{+0.08}_{-0.07}$, also slightly higher than the inner stellar halo, but again with overlapping error bars.

\subsection{NGC 891}
The panoramic view of NGC 891 presented in \cite{Mouhcine10} shows abundant stellar halo substructure (very apparent in Fig.\ \ref{substructuremap}), most notably a large tidal stream that appears to wrap around the galaxy more than once. The GHOSTS HST fields are oriented on the side of NGC 891 with more limited coverage, but it is clear that Fields 5, 6, and 7 overlap with overdensities associated with NGC 891's tidal stream in both Fig.\ \ref{substructuremap} and from the flattening of the surface density profile at those radii in \citet{Harmsen17}. The AGB/RGB ratio for this substructure region we found to be $-0.99^{+0.05}_{-0.04}$, similar to the major axis value of $-0.96\pm0.04$. Meanwhile, the inner part of the stellar halo gives an AGB/RGB ratio of $-1.22^{+0.05}_{-0.07}$, which is significantly lower than the other two.

\subsection{M81 (NGC 3031)}
M81 (NGC 3031) shows much higher uncertainties in the AGB/RGB ratio compared to the other three galaxies due to a significantly smaller number of both AGB and RGB stars, which can be seen in the stacked CMDs in figure \ref{Composite CMDs}. In spite of the large error bars we are able to obtain ratios for the {major axis$+$M82 halo fields and the minor axis fields.} The {major axis$+$M82 halo} ratio is larger than the minor axis (but with little significance, as the error bars largely overlap), with the {major axis$+$M82 halo} having a ratio of $-1.10^{+0.20}_{-0.24}$ compared to the minor axis ratio of $-1.65^{+0.21}_{-0.41}$.  

\subsection{General Trends}

While the three galaxies clearly differ in their properties, there are some general trends that are hinted at by the data. Firstly, bright AGB stars exist in all three stellar haloes, indicating that all haloes have contributions from non-ancient stellar populations. There is a general tendency towards the minor axis inner halo to have the lowest AGB/RGB ratio --- this is seen in all galaxies, but in some cases with very limited significance. Substructure fields have higher AGB/RGB ratios, generally consistent (in the case of NGC 253 and NGC 891 where substructure is distinct from the major axis) with the AGB/RGB measurements of the major axis fields.

\section{Interpretation and Discussion} 
\label{Interpretation}

In the previous section, we saw that the stellar haloes of NGC 253, NGC 891 and M81  lack significant main sequence or helium-burning populations, suggesting a low to negligible present-day star formation rate. In contrast, the RGB and AGB populations imply considerable star formation at earlier times, extending towards intermediate times in the case of NGC 253 and NGC 891, of order several Gyr ago.  Furthermore, substructure shows higher AGB/RGB ratios, implying that star formation continued until more recently. In this section, we will explore these results in more depth. We first attempt to constrain the likely relationship between AGB/RGB ratio and SFH using both simple modeling and real galaxies with published SFHs for which we can make AGB/RGB measurements. As we argued earlier, the accretion and tidal disruption of a satellite ultimately halts its star formation
(through ram-pressure, tidal effects, or a combination of both). Accordingly, we focus on exploring and quantifying the relationship between AGB/RGB ratio and when star formation shuts off in a stellar population; we find a tight relationship, offering a useful diagnostic of star formation shut-off time in satellites and stellar haloes. We then briefly discuss the possible implications of the differences in AGB/RGB ratio within haloes, and conclude with a comparison between inferences from our measurements and hydrodynamical models of galaxy formation.

\subsection{AGB/RGB ratios from stellar population synthesis models.} \label{subsec:ratio_mod}

\begin{figure*}\centering
\includegraphics[width=175mm]{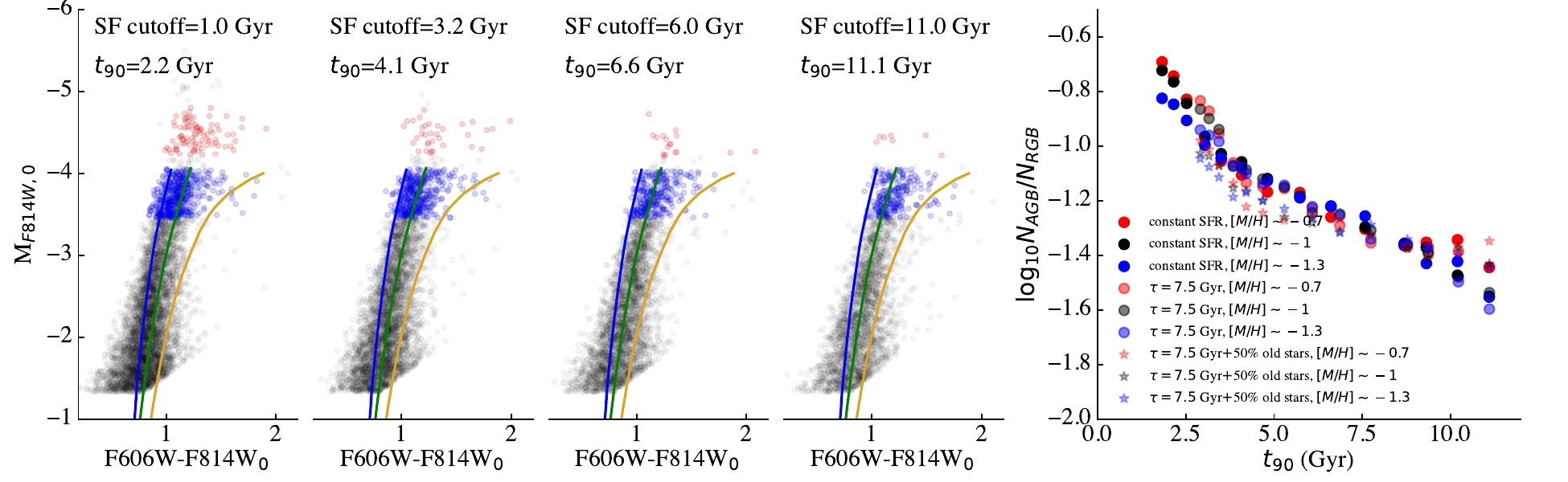}
\caption{A visualization of model CMDs with different star formation (SF) cutoff times. The left four panels show a mock CMD for a population with constant SFH before its SF is truncated between 1 and 11\,Gyr ago, with a Gaussian metallicity distribution of width 0.3\,dex centered on $[M/H] \sim -1$, with stars that would be selected by our RGB and AGB cuts highlighted in blue and red respectively. Three RGB isochrones are shown for illustrative purposes, at $[M/H] = -1.5, -1, -0.5$ {for an age of 8\,Gyr}, from left to right (blue, green, orange). Each of these model CMDs give AGB/RGB ratios which we can then compare with observations. In the right-most panel, we quantify a strong relationship between $t_{90}$ and $\log_{10}(N_{AGB}/N_{RGB})$, and that this relationship is not particularly sensitive to modest variations in metallicity around $[M/H] \sim -1$ and star formation history before the cutoff ({see text for details}).}
\label{AGB RGB Models}
\end{figure*}

We first explore model expectations for a relationship between AGB/RGB ratio and the SFH, with a particular focus on comparing with measures of when star formation halts in a stellar population.  
We show the results of our CMD modeling in Fig.\ \ref{AGB RGB Models}, using {Padova isochrones, {CMD version 3.2} with PARSEC evolutionary tracks version 1.2S and COLIBRI isochrones for the thermally-pulsing AGB star phase following \citet{Marigo2017}; see Appendix \ref{ap:agb_unc} for more discussion of the impacts of model uncertainties}. We explore a range of metallicities centered broadly around $[M/H]\sim -1$ (matching the observed metallicities of the stellar haloes of M81, NGC 253 and NGC 891; \citealt{Monachesi16a}), and adopt a Gaussian metallicity distribution with a dispersion 0.3\,dex in all cases. While our results hold for a variety of different SFHs, for illustrative purposes, we choose to show two SFHs before truncation --- constant SFH starting 12\,Gyr ago, or a SFR that declines from 12\,Gyr ago towards the present day with an exponential decay time-scale of $\tau = 7.5$\,Gyr. 

In order to parameterize the SF truncation time in a way that connects with observations, we follow \citet{Weisz2015} in adopting the time at which 90\% of the SF in a galaxy is complete, $t_{90}$. Not only is this measure particularly suitable for AGB stars --- which are insensitive to the SFH in the last few hundred Myr --- but also is a reasonably robust measure of SFH. As \citet{Weisz2015} discuss, blue stragglers --- intermediate-mass stars that were likely formed in stellar mergers of lower-mass stars ---  and e.g., occasional bright blue foreground stars that are not part of the galaxy can drive SF history fitting algorithms to require a small amount of recent star formation to fit an otherwise older population with no ongoing SF. The parameter $t_{90}$, because it neglects the most recent 10\% of the SFH, is insensitive to this low-level spurious SF, with the disadvantage that for a system with actual ongoing recent SF, $t_{90}$ misses the last 10\% of {star formation, setting a floor of $t_{90}\sim 1$\,Gyr for systems with ongoing SF}. In Fig.\ \ref{AGB RGB Models}, we plot our AGB/RGB ratios as a function of $t_{90}$ to place it on a scale that can be observed, noting that very similar trends are seen as a function of shutoff time. 

The left 4 panels show mock CMDs with $[M/H]\sim -1$ and a constant SFH; the right-most panel shows how the AGB/RGB ratio varies including a wider variety of metallicities and SFHs {(including an exponentially-declining star formation rate with exponential fall-off time $\tau=7.5$\,Gyr, and this declining model with half of the stellar mass being in an old population between 8 and 12 Gyr old)}. The AGB/RGB ratio reflects when SF shuts off in a stellar population in a way that is largely insensitive to the pre-truncation SFH and metallicity, for the metallicities around $[M/H]\sim -1$ relevant for the stellar haloes of NGC 253, NGC 891 and M81 \citep{Monachesi16a}. {We caution against over-interpreting the tightness of this relation --- while models agree that AGB/RGB ratio should be lower for old populations than $\sim$Gyr-old populations, specifics about how the thermally-pulsing AGB phase is modelled can dramatically depress the population of bright AGB stars for old ages (see Appendix \ref{ap:agb_unc} or e.g., \citealt{Marigo2017} or \citealt{Pastorelli2019}).}

\subsection{AGB/RGB ratio from observations of Local Group satellites} \label{subsec:ratio_obs}

\begin{figure*}\centering
\includegraphics[width=175mm]{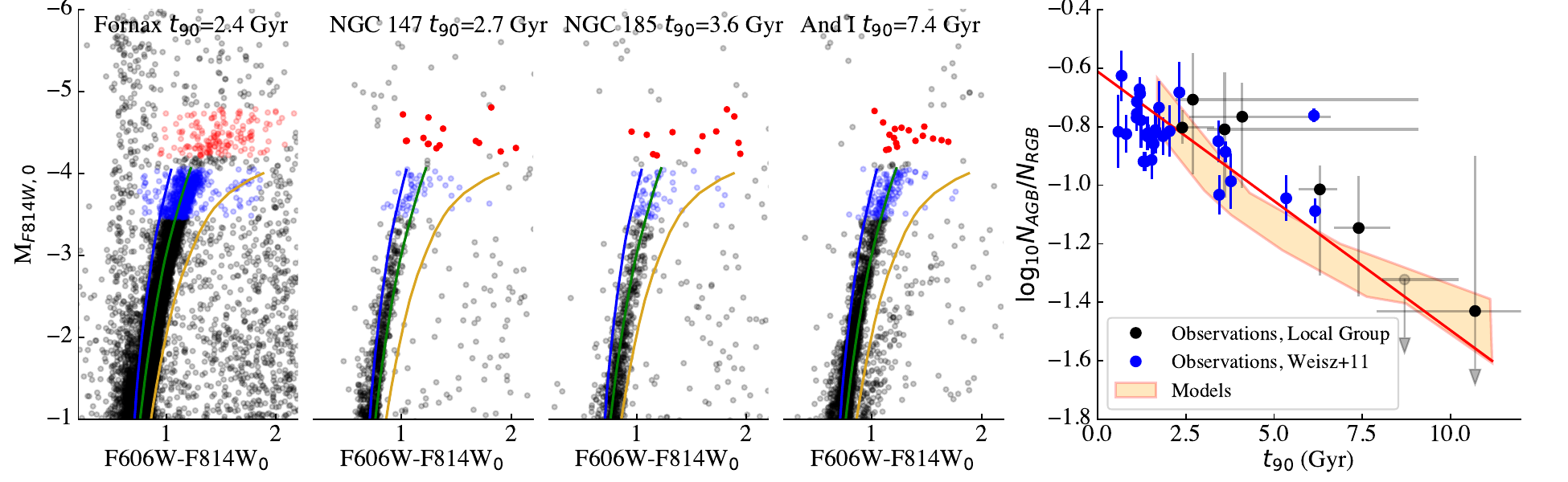}
\caption{Left four panels: Observed Local Group dwarf galaxy CMDs, arranged by inferred $t_{90}$. Stars that would be selected by our RGB and AGB cuts are highlighted in blue and red respectively. Again, three RGB isochrones are shown for illustrative purposes, at $[M/H] = -1.5, -1, -0.5$ {for an age of 8\,Gyr}, from left to right (blue, green, orange). In the right-most panel, we show the observed values of AGB/RGB ratio as a function of $t_{90}$ for galaxies from the Local Group (in black with upper limits in gray) and from the Local Volume (in blue). An orange region roughly spans the relationship shown by models from Fig.\ \protect\ref{AGB RGB Models}. The red line shows the best-fitting model (linear fit plus intrinsic scatter) of $t_{90}$ as a function of $\log_{10} (N_{AGB}/N_{RGB})$ which we use to estimate $t_{90}$ values on the basis of our measurements of AGB/RGB ratio.}
\label{AGB RGB Observations}
\end{figure*}

{Fig.\ \ref{AGB RGB Models} indicates that AGB/RGB ratio should be sensitve to age, and that $t_{90}$ is a reasonable way to parameterize the aspects of star formation history reflected in an AGB/RGB ratio. Yet, given the model uncertainties (Appendix \ref{ap:agb_unc}), it is imperative to calibrate this relationship using observational data. } We explore this relationship using two sets of galaxies.

Our first sample is satellites in the Local Group that are close enough that accurate SF histories exist have been inferred from deep CMDs \citep{Weisz2015,Weisz2019}, and have enough AGB stars to yield a measurement or constraining limit on the AGB/RGB ratio (limiting us to relatively massive satellites, following \S \ref{aggregation}; {see Appendix \ref{ap:obscal}}). 
{We show the resulting AGB/RGB ratios and $t_{90}$ values} in Fig.\ \ref{AGB RGB Observations}. Like Fig.\ \ref{AGB RGB Models}, four representative CMDs are shown, ordered by $t_{90}$ in the four left-most panels. In the right-most panel, the observed AGB/RGB star ratio for Local Group satellites is shown as a function of observed $t_{90}$ (in black, with upper limits in gray). The orange region illustrates the range of AGB/RGB ratios as a function of $t_{90}$ spanned by the same models presented in Fig.\ \ref{AGB RGB Models}. 

{Because there} are relatively few galaxies with the combination of sufficient luminosity to have a sizeable AGB star population and published SFH information, {we} have augmented the Local Group sample with those Local Volume galaxies with published SFH information from \citet{Weisz2008} and \citet{Weisz2011} and published ACS photometry in any two of the F475W, F555W, F606W and F814W filters from \citet{Dalcanton09}; {see Appendix \ref{ap:obscal}}. These points are shown in blue in Fig.\ \ref{AGB RGB Observations}. These galaxies tend towards much lower values of $t_{90}$ and higher AGB/RGB ratios than the Local Group sample, populating that part of the relation but not adding to the set of galaxies with older $t_{90} > 6$\,Gyr. {In particular, many of the systems with $t_{90}<1.5$\,Gyr have some degree of ongoing star formation \citep{Weisz2011}, in agreement with model expectations of $t_{90}\sim1$\,Gyr or less for such systems. } 

While the uncertainties are considerable, the observations follow a similar trend to the models, but with more scatter than the models would naively indicate, with factors of five or more variation in AGB/RGB ratio correlating with inferred $t_{90}$. This scatter may stem from unrecognised uncertainties in AGB/RGB ratio estimation, reflect SFH uncertainties or limitations in how $t_{90}$ parameterizes the SFH features most important for setting AGB/RGB ratio, or may reflect modest shortcomings in the (challenging) modeling of the lifetimes, colours and magnitudes of the extremely bright and short-lived stars that we investigate. We choose to parameterize the variation of $t_{90}$ as a function of observed $\log_{10} (N_{AGB}/N_{RGB})$ by fitting a straight line (where the $\log_{10} (N_{AGB}/N_{RGB})$ is taken as the independent variable), shown in red in Fig.\ \ref{AGB RGB Observations}. We assume that the $t_{90}$ is drawn from a Gaussian distribution function around $t_{90,expect}$:
\begin{equation}
    P(t_{90}) = \frac{1}{\sqrt{2 \pi \sigma_t}} e^{-\frac{(t_{90}-t_{90,expect})^2}{2 \sigma_t^2}},
\end{equation}
where $\sigma_t$ is the intrinsic scatter of $t_{90}$ for a given observed AGB/RGB ratio. The value $t_{90,expect}$ is assumed to vary linearly with $\log_{10} (N_{AGB}/N_{RGB})$:
\begin{equation} \label{eq:t90}
    t_{90,expect} = t_{90,-1} + \alpha (\log_{10} (N_{AGB}/N_{RGB}) + 1), 
\end{equation}
where $t_{90,-1}$ is the expected $t_{90}$ at $\log_{10} (N_{AGB}/N_{RGB})=-1$, and $\alpha$ is the slope of the relation. We fit this using Markov Chain Monte Carlo fitting using uninformative priors, finding 68\% confidence intervals of $t_{90,-1} = 4.4 \pm 0.3$, $\alpha = -11.3\pm1.6$, and $\sigma_t = 1.45\pm0.18$\,Gyr.
In what follows, we will base inferences about stellar halo SFHs on this fitted relation between $t_{90}$ and $\log_{10} (N_{AGB}/N_{RGB})$, including the contribution of the intrinsic/unmodeled scatter $\sigma_t$ in quadrature.

\subsection{Older well-mixed haloes and younger substructure}

With insight from models and observations from Local Group galaxies in hand, we can now start to interpret the AGB/RGB ratios in the stellar haloes of NGC 253, NGC 891 and M81, and explore the implications of the spatial variations that we see in AGB/RGB ratio in those galaxies. Observed values of AGB/RGB ratio were used to inform $t_{90}$, sampling from the posterior distribution of the best-fitting $\log_{10} (N_{AGB}/N_{RGB})$ vs.\ $t_{90}$ relation, accounting for $\sigma_t$ by adding it in quadrature to the inferred error bars. 
We present inferred $t_{90}$ values and their 68\% confidence intervals in Table \ref{AGB RGB Ratios Table}. We present values for the well-mixed inner haloes of the three galaxies and the major axes of each galaxy. The major axis {fields of M81 also contain} tidal debris from M82's ongoing disruption, and the fields within which we can measure AGB/RGB ratio are dominated by that tidal debris (see also \citealt{Harmsen17} and \citealt{Smercina20}). We also present inferred $t_{90}$ values for NGC 253's 30\,kpc shell \citep{Greggio14} and NGC 891's stellar stream \citep{Mouhcine10}.

Turning first to the minor axis inner stellar halo measurements (which we take to represent the well-mixed parts of their haloes), we find that the AGB/RGB ratios (and inferred $t_{90}$ values) of the inner haloes are too high to be consistent with {a solely ancient stellar population} for NGC 253 and NGC 891 (6.2\,Gyr and 6.9\,Gyr, with around 1.5\,Gyr uncertainty on each value; Table \ref{AGB RGB Ratios Table}). M81 has weaker constraints, with a poorly-measured age $11.8\pm4.0$\,Gyr that is  consistent with ancient stellar populations. These values suggest growth of NGC 253's and NGC 891's well-mixed halo components until at least $z \sim 1$, $\sim 7$\,Gyr ago.

M81, in contrast, has a poorly-measured AGB/RGB ratio on its well-mixed minor axis implying that it has $t_{90}>8$\,Gyr. M81's halo has an existing age measurement in a single deep field (\citealt{Durrell10}; see \citealt{Williams09} for a deep SFH measurement of M81's outer disc), where a mean age of $9\pm2$\,Gyr was derived from the colour of the RGB, the brightness of the RGB bump, and the brightness of the red clump. This independent measure of mean age lends credibility to the less precise AGB/RGB-derived estimate of $t_{90}$. Interestingly, M81 joins the Milky Way as a galaxy whose dominant accretion event, giving rise to its well-mixed stellar halo, appears to be relatively ancient (e.g., \citealt{Gallart2019}). 

Yet, in all three galaxies, there are regions of the halo with higher AGB/RGB ratio that indicate that important amounts of accretion have continued well after that time. The AGB/RGB ratios $\log_{10} (N_{AGB}/N_{RGB}) \sim -1.04^{+0.11}_{-0.20}$ in NGC 253's overdense shell region \citep{Greggio14} and major axis fields suggest $t_{90}$ values of roughly 5\,Gyr ago, rather younger than NGC 253's well-mixed halo. 
NGC 891's prominent tidal streams \citep{Mouhcine10} and major axis have $\log_{10} (N_{AGB}/N_{RGB}) \sim -1.0$ indicating $t_{90} \sim 4$\,Gyr. In NGC 891's case, the RGB map from \citet{Mouhcine10}, shown also in Fig.\ \ref{substructuremap}, suggests that the major axis GHOSTS fields appear to overlap with debris/wraps of the tidal streams. In the case of M81, the tidal debris from M82's accretion (see e.g., \citealt{Okamoto15,Smercina20}) along M81's major axis has a relatively low AGB/RGB ratio, $\log_{10} (N_{AGB}/N_{RGB}) = -1.10^{+0.20}_{-0.24}$, corresponding to $t_{90} \sim 5$\,Gyr. This is younger than the rest of M81's halo.

Qualitatively, these measurements demonstrate that {\it dynamically-young} debris from more recent accretions that has not had time to phase-mix into a smoother halo also show signs of {\it younger stellar populations}. Quantitatively, $t_{90}$ has complexities in its interpretation. At the very least, Fig.\ \ref{AGB RGB Observations} shows  $t_{90} \sim 1$\,Gyr for systems with {\it ongoing} SF (as expected); $t_{90}$ is always an upper limit to when SF shuts off in a system. There are other, more difficult to circumvent challenges, however. For example --- our $t_{90}$ measurements for the M81 group major axis {fields} reflect the star formation history of the (previously) very outer parts of M82, and includes the impacts of pre-existing age gradients and imprints of the time at which star formation ceased as the stars were pulled from M82. Similar issues may complicate the quantitative interpretation of $t_{90}$ estimates for NGC 253's shell and NGC 891's streams.

\subsection{Comparison with stellar haloes from cosmologically-motivated models of galaxy formation} \label{subsec:TNG}

\begin{figure}\centering
\includegraphics[width=80mm]{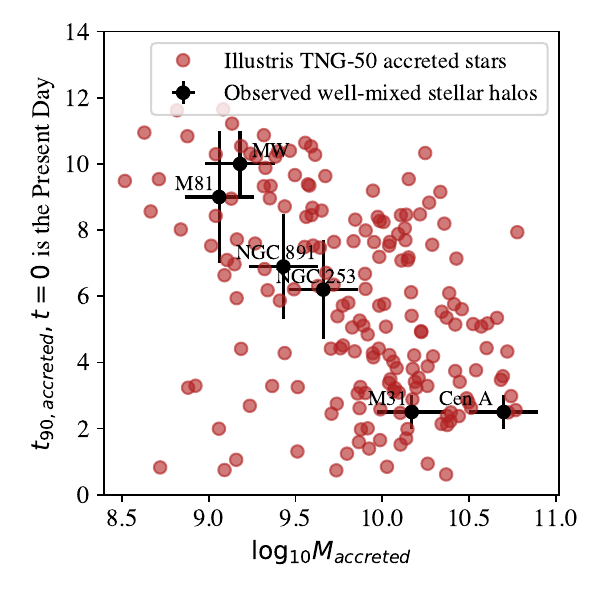}
\caption{The $t_{90}$ of the accreted component of Milky Way-mass galaxies from the TNG-50 simulation is shown as a function of their total accreted mass (red). Overlaid in black are estimates of $t_{90}$ for six observed systems: (left to right) M81, the Milky Way, NGC 891, NGC 253, M31 and Cen A.  }
\label{t90_models_obs}
\end{figure}

This work provides two additional constraints on the star formation history of the stellar haloes of Milky Way-mass galaxies in the Local Volume, bringing the number of galaxies with useful constraints on when star formation stopped in their haloes to six: (left to right in Fig.\ \ref{t90_models_obs}) M81 \citep{Durrell10}, the MW \citep{Gallart2019}, NGC 891 and NGC 253 (this work). In the case of M31 we adopt a $t_{90} = 2.5\pm0.5$\,Gyr given that the last significant episode of inner stellar halo star formation spans 2--3\,Gyr ago (\citealt{Brown06}, as visualised in Fig.\ 1c of \citealt{DSouza18b}). We include a $t_{90} = 2.5\pm0.5$\,Gyr estimate for the stellar halo of NGC 5128 (Centaurus A) --- an elliptical galaxy --- given the claim from \citet{Rejkuba2022} of a significant mass of stars formed 2-3\,Gyr ago from both AGB stars within $\sim30$\,kpc (using a similar method to our own; \citealt{Rejkuba2022}) and CMD modeling of a deep outer field ($R_{projected} = 38$\,kpc) of NGC 5128 \citep{Rejkuba2011}. These fields are sufficiently distant from the centre of Cen A that both \citet{Rejkuba2011} and we assume that these halo populations primarily reflect the SFH of the Cen A's accreted stellar population. We show these estimates in Fig.\ \ref{t90_models_obs} as a function of total stellar halo mass (from \citealt{Harmsen17} for M81, NGC 891 and NGC 253; from \citealt{Deason2019} for the MW; from \citealt{DSouza18b} for M31, and from \citealt{DSouza18a} for NGC 5128). 

It is clear that the $t_{90}$ values for this small observed sample of haloes tends towards younger ages for higher stellar halo masses. Pearson's correlation coefficient for these six galaxies is $r=-0.94$, corresponding to a 0.5\% chance of these six points being drawn from an uncorrelated parent dataset by chance alone.  While important amounts of late-time halo formation and assembly do not fit well with extrapolations of the intuition developed with our own stellar halo \citep[e.g.,][]{Kalirai2012,Gallart2019}, ongoing assembly of haloes, particularly massive ones, is an expectation of cosmological models of galaxy formation \citep{DSouza18a,DSouza18b}. We choose to illustrate this by measuring $t_{90}$ values for the accreted components of MW-mass simulated galaxies from the TNG-50 simulation. 

TNG-50 \citep{Pillepich2018,Pillepich2019}
simulates a large cosmological volume (51.7 Mpc on a side) with high
resolution (300\,pc softening length for the collisionless particles), enabling the analysis of the detailed properties of accreted stellar haloes. 
We selected 183 systems 
with stellar masses 3$\times 10^{10} M_{\odot}$ to 15$\times 10^{10} M_{\odot}$ and dark matter halo masses between 0.5-3$\times 10^{12} M_{\odot}$ without regard to galaxy morphology --- we note that the results are largely unchanged if we require that the MW-mass system has significant rotation to focus on systems more likely to have galactic discs.  
%We choose galaxies that have significant rotation by requiring that more than 40\% of stars are on orbits that have a circularity $\epsilon = J_z/J(E) > 0.7$, where $J_z$ is the specific angular momentum of a particle around the angular momentum axis of the stellar body of a galaxy, and $J(E)$ is the maximum angular momentum of the 100 particles with the most similar total binding energies (see also \citealt{Genel2015}). 
Stellar particles born on the main progenitor branch subhalo are labeled {\it in-situ}; particles born outside are labeled accreted. We calculate $t_{90}$ directly from the cumulative star formation history of the accreted particles\footnote{While it is possible to attempt to create mock AGB/RGB measurements for inner stellar halo fields from stellar population synthesis models, doing so is subject to additional simulation (e.g., metallicity) and stellar population synthesis systematic concerns (e.g., AGB lifetimes, magnitudes and colours; {Appendix \ref{ap:agb_unc}}), and doing so is beyond the scope of this work.}. We show the accreted masses and $t_{90}$ values for these accreted haloes in Fig.\ \ref{t90_models_obs} as red symbols. 

TNG-50 stellar haloes show a wide range of $t_{90}$ values, with a tendency towards younger $t_{90}$ values for more massive haloes. While there is considerable scatter in this relationship (Pearson's $r=-0.40$), the chance of this relationship being drawn from an uncorrelated dataset is $\sim10^{-8}$. This trend, in a broad sense, is a direct consequence of hierarchical galaxy formation. If a galaxy is to have a large stellar halo, it needs to have accreted a satellite with large stellar mass. Since we are considering a relatively narrow range of total dark matter halo masses when we consider Milky Way-mass galaxies, this limits the size of the dark matter halo that the galaxies can merge with. Considering relatively large secondaries (in dark matter mass), early mergers would have little time to form stars in its halo, favoring the formation of a lower mass stellar halo. Merging later allows much more time for the secondary to form stars, driving the formation of a larger halo with a later $t_{90}$. This overall trend is in accord with observations; what remains to be seen is if the considerable scatter in halo $t_{90}$ values at a given stellar halo mass for the the 183 simulated systems is simply not yet evident with only six observed haloes, or is a consequence of some limitation in the TNG-50 simulations.  

\subsection{Limitations of this work}

{While the work described in this paper describes a practical path forwards in placing useful age constraints on nearby galactic stellar halos, there are a number of limitations that it is important to consider. }

{Both the choice of $t_{90}$ as a measure of star formation history and its interpretation as a halo star formation shut-off time are central to our work. The star formation histories of the MW's (e.g., \citealt{Gallart2019}) and M31's (e.g., \citealt{Brown06,Brown08}) stellar halos both show a dramatic decrease in SF at the time at which its most important merger took place; models show similar behaviour in their halo star formation histories (e.g., \citealt{DSouza18b}, \citealt{Monachesi19}). Many still-existing satellites show evidence of a decrease in star formation  
\citep[e.g.,][]{Slater2014_sat,Fillingham2015,Fillingham2016}, a behaviour also seen in hydrodynamical models \citep{Samuel2022}. Yet, it is similarly clear that superimposed on this broad behaviour will be complications from the richness of gas physics in galaxy group interactions. One need look no further than the M81 group to see this. The M81 group has a widespread {\sc Hi} tidal debris field \citep[e.g.,][]{Yun99,deBlok2018} which is forming stars \citep{Okamoto15,Okamoto2019} --- a clear case of {\it in-situ} halo star formation. M82, its largest satellite, is starbursting, and many members of its group have significant star formation \citep{Weisz2008}. Beyond this, there are claims that star formation in satellites around MW-mass galaxies is more prevalent than would be suggested by the Local Group \citep{SAGAII,Carlsten22} and found in hydrodynamical models \citep{Font22}. This does not invalidate the core assumption that stellar halo star formation histories carry information about merger histories, but it does emphasize that a more precise understanding will require that the community builds a deeper understanding of group-wide star formation in galaxy groups, particularly interacting groups.        }

{Another key complication to bear in mind is stellar halo substructure. For MW-mass galaxies, all stellar halos are contributed to by many satellites, most of which are not fully phase-mixed \citep{Bullock01,BJ05,Cooper10,Monachesi19}. Indeed, the Milky Way's and M31's stellar halos both show contributions from numerous satellites \citep[for relatively recent discussions of this issue see e.g.,][]{McConnachie18,Malhan22}, and both NGC 253 \citep{MalinHadley97,Bailin11,Greggio14} and NGC 891 \citep{Mouhcine10} show clear substructure from recent accretion. This is an inevitable feature of halos that presents not just observational challenges, but requires that we choose how we think about stellar halos to focus on features of their growth histories that are more amenable to measurement and more impactful in shaping the main galaxies. Most stellar halos are predicted to have had one accretion that dominates the halo properties --- their mass, metallicity, and, critically for our work, star formation history \citep[e.g.,][]{Cooper10,Deason16,Harmsen17,DSouza18a}. While the rarity of AGB stars necessitated that large regions were aggregated to constrain $t_{90}$, we attempted to be sensitive to the potential importance of substructure by choosing our regions for study based on previous signs of substructure from large-scale imaging surveys (M82's halo/tidal debris, NGC 253's shells and NGC 891's streams), in addition to a minor-axis inner halo measurement that is hoped to reflect more of the phase-mixed debris, dominated by the largest merger experienced by the galaxy. Yet, it is clear that many satellites contributed to the regions that we measured, and this remains an important and inevitable limitation of this work. It will be useful to explore this issue more explicitly both theoretically and observationally, in the observational case particularly in M31 where deep SFH measurements in various halo fields exist; indeed, this may motivate wide-field space-based maps with e.g., the Nancy Grace Roman Space Telescope. }

{This challenge is exacerbated by the rarity of AGB stars;} $\sim10^8 M_{\odot}$ of stars needs to be surveyed to yield the $\sim 100$ AGB star sample that permits accurate measurement of the AGB/RGB ratio. Our best measurements are for NGC 891, which has the triple advantage of distance (one HST field covers nearly seven times the area of each field in NGC 253 or M81), a significant stellar halo mass ($\sim 3\times10^9 M_{\odot}$; \citealt{Harmsen17}), and a relatively prominent bright AGB star population. In contrast, our worst measurements, for M81, reflect its nearby distance, modest stellar halo mass of $\sim 10^9 M_{\odot}$ \citep{Harmsen17,Smercina20} and older age \citep{Durrell10}. For nearby systems, wide-field space-based imaging from the {\it Nancy Grace Roman Space Telescope} will likely be necessary to survey sufficient areas with the required sensitivity. For more distant systems, the modest field sizes offered by {\it HST} or {\it JWST} will likely be sufficient.  

A minor limitation is the red extent of our selection boxes; the minor axis CMDs (Fig.\ \ref{Composite CMDs}) are populated enough that modest metal-rich tails in the RGB populations are visible. We have confirmed that including these regions in our analysis affected our inferences by less than their quoted uncertainties, largely owing to the relatively small numbers of stars in these parts of the CMD. Indeed, to measure a $t_{90}$ value, it may not be necessary to include high or low metallicity tails in the selection regions, as the $t_{90}$ value for the bulk of the stellar population should be shared also by those stars in the sparsely-populated tails of the distribution. 

One fundamentally limiting choice that we made is to focus entirely on AGB/RGB ratio as a measure of SFH. By its nature, the AGB/RGB ratio is sensitive to both the numerator (the number of intermediate-age AGB stars) and the denominator (a broad measure of the total mass formed, although also weighted towards more recent star formation). In some sense, the AGB/RGB ratio can be thought of as a specific star formation rate indicator, focused on intermediate ages instead of present-day SFR. Indeed, the AGB/RGB ratio clearly varies (at the order of magnitude level or more) with the SFH (Figs.\ \ref{AGB RGB Models} and \ref{AGB RGB Observations}), as parameterized by $t_{90}$. {That the observations show a scattered relation, despite the variations in SFH within the calibrating galaxy dataset  (Fig.\ \ref{AGB RGB Observations}), suggests that AGB/RGB ratio will be useful at least at the level of $\pm2\,$Gyr for constraining $t_{90}$. Yet, the sensitivity of AGB/RGB ratio to model parameterizations and ingredients (Appendix \ref{ap:agb_unc}; \citealt{Marigo2017,Pastorelli2019}) highlights the importance of further model work and observational calibration if one wishes to refine AGB-based measures of SFH. }

Yet, the choice to use {colour--magnitude diagram} selection regions {\it at all} for study is a clear limitation. In our current work, we deliberately avoided attempting to model the full distribution of AGB magnitudes and colours, in part owing to our concerns about the difficulty of modeling the bright thermally-pulsing AGB stars that are being analyzed here \citep[e.g.,][]{Marigo2017}. 
However, the isochrones in Fig.\ \ref{AGBRGB_iso} and the distribution of predicted AGB stars in Fig.\ \ref{AGB RGB Models} show the model expectation that the brightest AGB stars will only arise in important numbers for more recent star formation. For example, in the well-populated minor axis CMDs for NGC 253 (Figs.\ \ref{AGBRGB_iso} and \ref{Composite CMDs}) and NGC 891 (Fig.\ \ref{Composite CMDs}), there is a relatively sharp feature in the CMD at the bright end of the AGB, at $m_{\rm F814W} \sim 23.2$ for NGC 253 and $m_{\rm F814W} \sim 24.8$ for NGC 891. Brighter than this cut, there are relatively few stars; fainter than this cut there are many more AGB stars, possibly reflecting in a more detailed way the time at which star formation shut off in their haloes. This opens the possibility that the AGB luminosity function may be a promising next angle to explore to generate more detailed information about the time at which star formation shuts off in a halo (e.g., \citealt{Rejkuba2022}), {if models and observational calibration efforts \citep{Marigo2017,Pastorelli2019} mature enough over the next few years.  }

\subsection{Discussion and Outlook}

Recalling the difficulty of measuring stellar halo SFHs outside of the Local Group, this work is notable for increasing the number of haloes with useful SFH constraints by 50\%, from four (MW, M31, M81 and Cen A) to six (including NGC 253 and NGC 891). We highlighted the potential utility of AGB/RGB ratios in constraining the commonly-used $t_{90}$ measure of when the bulk of star formation has ceased in a stellar halo, giving a measure of when stellar halo assembly was largely finished. 

With these measurements in hand, our data reveal a clear but poorly-populated anticorrelation between the mass of a stellar halo and its inferred $t_{90}$ value. While this runs counter to decades of experience honed in the Milky Way's old, metal-poor halo \citep[e.g.,][]{Kalirai2012,Gallart2019}, simulations of stellar halo growth in Milky Way-mass galaxies by merging and accretion have long predicted that there should be a wide range in stellar halo mass, and that the most massive haloes built up much more recently than their low-mass cousins \citep[e.g.,][]{Purcell07,Cooper10,Deason16,Amorisco2017,DSouza18a}. This correlation was hinted at with the realization that M31's and Cen A's stellar haloes have a substantial population of intermediate-age stars \citep[e.g.,][]{Brown06,Brown08,Bernard2015,DSouza18b,Rejkuba2011,Rejkuba2022}. In adding more galaxies to the sample, and comparing explicitly with simulations, this work more clearly confirms this theoretical expectation, and more clearly delineates how stellar haloes might provide powerful constraints on a galaxy's merger history. 

Because the AGB/RGB ratio is measured using the brightest AGB and RGB stars (our work and \citealt{Rejkuba2022}), it is an unusually observationally-accessible indicator of $t_{90}$. Recalling the necessity to survey large areas to assemble large enough samples of luminous AGB stars, this technique is currently most applicable for systems towards the far edge of the Local Volume using {\it HST} data (more distant than e.g., 7-9\,Mpc) or beyond (in the JWST era, where its larger aperture makes it feasible to study the haloes of more distant galaxies, at $D>10$\,Mpc). 

{Acknowledging the critical uncertaines at the level of AGB model calibration (Appendix \ref{ap:agb_unc}), if the calibration effort illustrated in Fig.\ \ref{AGB RGB Observations} (very similiar in many respects to that of \citealt{Rosenfield2016}) can be taken at face value}, AGB/RGB ratio measurements will be able to place competitive bounds on stellar halo $t_{90}$, not requiring any additional data over and above those required to constrain the stellar halo mass \citep[e.g.,][]{Harmsen17,InSung2020M101} and metallicity \citep[e.g.,][]{Monachesi13,Rejkuba14,Monachesi16a}. Given that the required measurements are feasible within $<\sim$ 10 Mpc with HST and to larger distances with JWST, there is an opportunity to construct a sizeable, volume-limited sample of stellar haloes. Bearing in mind that $t_{90}$ will depend on the time of the last important merger, and the stellar halo mass and metallicity constrain the mass and metallicity of its most important merger partner \citep{Deason16,Harmsen17,DSouza18a,Monachesi19}, this sample would pave the way towards measuring the distribution of merger times and masses for Milky Way-mass galaxies \citep[e.g.,][]{Elais18,Sotillo22}, constraining how merger time and mass influence galaxy properties \citep[e.g.,][]{Bell17,Hammer2018,DSouza18b}, and exploring the importance of the group accretions for building up satellite populations \citep[e.g.,][]{Weisz2019,Patel2020,Dsouza2021,Smercina22,Bell22}. 

\section{Conclusions}

We have explored the bright AGB star content of the stellar haloes of three nearby highly-inclined disc galaxies with deep resolved-star HST observations --- NGC 253, NGC 891 and M81 --- in an effort to constrain their intermediate-age stellar populations. We use the ratio of bright AGB stars to the number of stars at the tip of the RGB to quantify the prominence of the bright AGB star population. All three stellar haloes have a bright AGB star population; it is the most prominent in the stellar haloes of NGC 253 and NGC 891. In all haloes, regions of the stellar halo along the major axis and fields with recognizable substructure (e.g., tidal streams or shells) have higher AGB/RGB ratio, albeit with poor number statistics in the case of NGC 253's and M81's haloes. 

In order to understand the implications of these measurements for our understanding of these galaxies' stellar haloes, we analyse stellar population synthesis models and a set of observed galaxies with full SFH information from full CMD modeling. For satellite-like SFHs which show ongoing SF, with a dramatic slow-down or shutoff (owing likely to a combination of tidal and ram-pressure effects), {models suggest and observations show} that the AGB/RGB ratio can give insight into the time before which 90\% of the mass in stars in a galaxy formed, $t_{90}$. We estimate that $t_{90}$ values for the well-mixed inner stellar haloes of $6.2\pm1.5$\,Gyr for NGC 253, $6.9\pm1.6$\,Gyr for NGC 891, and  $11.8\pm4.0$\,Gyr for M81, with younger values for their major axes and recognizable halo substructure. Combining these measurements with a previous measurement of M81's halo age (for a deep M81 halo pointing) or star formation shut-off times for the Milky Way, M31 and Cen A, we find a wide variation in stellar halo SFHs, with a tentative trend towards later star formation shut-off time for higher mass stellar haloes. Comparing these measurements with trends in $t_{90}$ values with accreted mass (broadly equivalent to stellar halo mass) for the accreted star particles from the TNG-50 hydrodynamical simulation for Milky Way like galaxies, we find good agreement --- this is a model prediction that has never before been tested. 

While it is likely that more information can be extracted from the bright AGB star population with different measurement techniques and {future} more model-informed analyses, this work highlights the promise of the bright AGB star population as a limited but, crucially, observationally-accessible measure of stellar halo star formation histories. Bright AGB stars are detected by the same resolved-star measurements that are required to infer stellar halo mass and metallicity from their RGB star content --- measurements that will be more straightforward with the JWST or the Nancy Grace Roman Space Telescope. Recognizing that these measurements constrain the most important mergers experienced by a galaxy, this offers the possibility of dramatic progress in the next years in understanding the long-term impact of mergers on the structures, star formation histories and satellite populations of a substantial, volume-limited sample of nearby galaxies.

\vspace{0.5cm}

This work was partly supported by HST grant GO-15230 provided by NASA through a grant from the Space Telescope Science Institute, which is operated by the Association of Universities for Research in Astronomy, Inc., under NASA contract NAS5-26555. We acknowledge support from the National Science Foundation through grant NSF-AST 2007065 and by the WFIRST Infrared Nearby Galaxies Survey (WINGS) collaboration through NASA grant NNG16PJ28C through subcontract from the University of Washington. AM gratefully acknowledges support by FONDECYT Regular grant 1212046 and by the ANID BASAL project FB210003, as well as funding from the Max Planck Society through a “PartnerGroup” grant. This research has made use of NASA's Astrophysics Data System Bibliographic Services. We made use of the following sofware: \texttt{Matplotlib} \citep{matplotlib}, \texttt{NumPy} \citep{numpy}, \texttt{Astropy} \citep{astropy}, \texttt{SciPy} \citep{scipy},  \texttt{Scikit-learn} \citep{scikit-learn}, and \texttt{emcee} \citep{emcee}.

\section{Data Availability}

The data underlying this article will be shared on reasonable request to the corresponding author.

\bibliographystyle{mnras}
%\interlinepenalty=10000
\bibliography{stellhalos}

\appendix

\section{The impact of stellar mass loss assumptions on the relationship between AGB/RGB ratio and $t_{90}$} \label{ap:agb_unc}

{Our paper focuses on using AGB stars brighter than the TRGB, detected in the optical. Such bright and optically-bright AGB stars are readily visible in relatively shallow CMDs such as our own, offering in principle a readily-accessible measure of SFH at intermediate ages. Yet, the AGB is one of the most uncertain phases of stellar evolution (see \citealt{Herwig2005} for a review). While we largely sidestep the impact of these theoretical uncertainties by calibrating the AGB/RGB ratio to $t_{90}$ using galaxies with observed AGB/RGB ratio and $t_{90}$ values, we consider it useful to briefly discuss the considerable challenges and uncertainties of modeling AGB stars, and calibrating these models, particularly for older stellar populations dominated by lower-mass stars.  

Different stellar population model frameworks use different approaches to model this exceptionally complex phase of stellar evolution. Models that integrate the full structure of the star during its evolution in the AGB phase (e.g., \citealt{Ventura2009,Cristallo2009}) are sufficiently computationally demanding that it is challenging to model the required diversity of AGB star parameters. One can circumvent that challenge using synthetic models that describe AGB star evolution using relations fitted to the results of full evolutionary models (e.g., \citealt{Cordier2007,Z2008}), with the acknowledgement that these are limited to the processes and parameter space explored by the full models. These approaches can be fruitfully combined, where full stellar structure modeling is reserved for the stellar envelope where it is most impactful, but model grids incorporating a wide range of stellar parameters and physical processes can be practically tackled (e.g., \citealt{Nanni2013,Marigo2013}). Furthermore, even within one approach, modeling choices make important differences to the final AGB star population (e.g., \citealt{Pastorelli2019}). As an illustration, the reader may wish to examine Fig.\ 11 of \citet{Marigo2017}, where those authors compare the evolutionary tracks from the COLIBRI code (incorporated into the PARSEC isochrones that we use in this work) with the BaSTI isochrones \citep{Cordier2007}, MIST isochrones \citep{Choi2016}, and a previous generation of their own work \citep{Marigo2008}. These models differ considerably in their predictions for the luminosities and temperatures (especially in the degree of variation), and in the prominence of the Carbon-rich AGB stars. While each model has its advantages and disadvantages, none are perfect and this underlines the important systematic uncertainties that will affect model-derived calibrations of AGB/RGB ratio as a function of SFH and metallicity. 

We choose to adopt the combination of the PARSEC \citep{Bressan2012} and COLIBRI \citep{Marigo2013} isochrones for our work.
We are sensitive primarily to AGB stars from relatively low-mass, older stellar populations. For these stars, mass loss both before and during the AGB phase and parameterization of the third dredge-up are particularly important for setting the lifetime of the thermally-pulsing AGB phase \citep{Rosenfield2014,Pastorelli2019}. This group (like other groups working in this field) has spent considerable effort on calibrating these features of their model using different datasets.  \citet{Rosenfield2014,Rosenfield2016} chose on one hand to calibrate the models using datasets very similar to those used in this paper, using SFHs recovered using full CMD modeling to predict AGB/RGB ratio. While many of their calibrating galaxies were rich in old stellar populations, the vast majority of the AGB stars were expected to be contributed by populations with ages $<6$\,Gyr (e.g., Fig.\ 12 of \citealt{Rosenfield2014} and Fig.\ 8 of \citealt{Rosenfield2016}), meaning that even calibrations that prioritize older populations cannot precisely pin down AGB lifetimes for the lowest mass stars. On the other hand, \citet{Pastorelli2019} chose to carefully analyse the SMC's AGB star content in light of information about its SFH from full CMD modeling. They showed that while the prescriptions of \citet{Rosenfield2016} matched the SMC's AGB star content in a broad sense (their \texttt{S\_00} model), they found that adjustments to mass loss (both before and during the AGB phase) and the modeling of the third dredge-up yielded much better matches to the SMC's AGB star population. Their \texttt{S\_35} model has been adopted as the default model for the new CMD 3.7 \footnote{\hyperlink{http://stev.oapd.inaf.it/cgi-bin/cmd}{http://stev.oapd.inaf.it/cgi-bin/cmd}} stellar population model interface in this metallicity range. 

\begin{figure}\centering
\includegraphics[width=70mm]{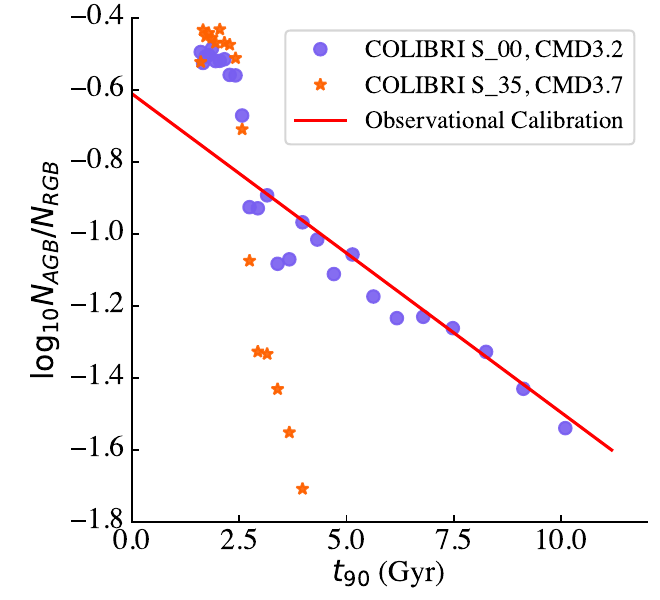}
\caption{We show predicted values of AGB/RGB ratio as a function of $t_{90}$ for two different PARSEC$+$COLIBRI stellar population models, illustrating the degree of variation between thoughtfully-calibrated stellar population models. The red line shows the linear best fit to the observational trend from Fig.\ \protect\ref{AGB RGB Observations}. }
\label{AGB RGB Diff Models}
\end{figure}

While a decrease in AGB/RGB ratio with increasing $t_{90}$ is a generic and inevitable outcome of stellar population modeling, the very significant differences between these parameterizations for low- mass stars with $M < 1.1M_{\odot}$ (see Fig.\ 14 of \citealt{Pastorelli2019}) drive the slope of the $\log AGB/RGB$--$t_{90}$ relation. In particular, model \texttt{S\_35} of \citet{Pastorelli2019} has very short (or zero) AGB lifetimes in this mass range, while the \texttt{S\_00} model of \citet{Rosenfield2016} has non-zero AGB star lifetimes to much lower stellar masses (therefore older population ages). The models that we show in Fig.\ \ref{AGB RGB Models} used CMD 3.2 PARSEC isochrones with the COLIBRI \texttt{S\_00} AGB tracks following \citep{Rosenfield2016}, and accordingly yield AGB stars even for older populations; we show them again in Fig.\ \ref{AGB RGB Diff Models}. Adopting instead CMD 3.7, PARSEC isochrones with the COLIBRI \texttt{S\_35} tracks (appropriate for this metallicity range) instead yields a much steeper fall-off in AGB/RGB ratio towards longer $t_{90}$ values, in contradiction to the observational constraints in Fig.\ \ref{AGB RGB Observations}, shown with a red line in Fig.\ \ref{AGB RGB Diff Models}.    } 

\section{Foreground and background contamination}
\label{ap:fgbgcont}

\begin{figure*}\centering
\includegraphics[width=140mm]{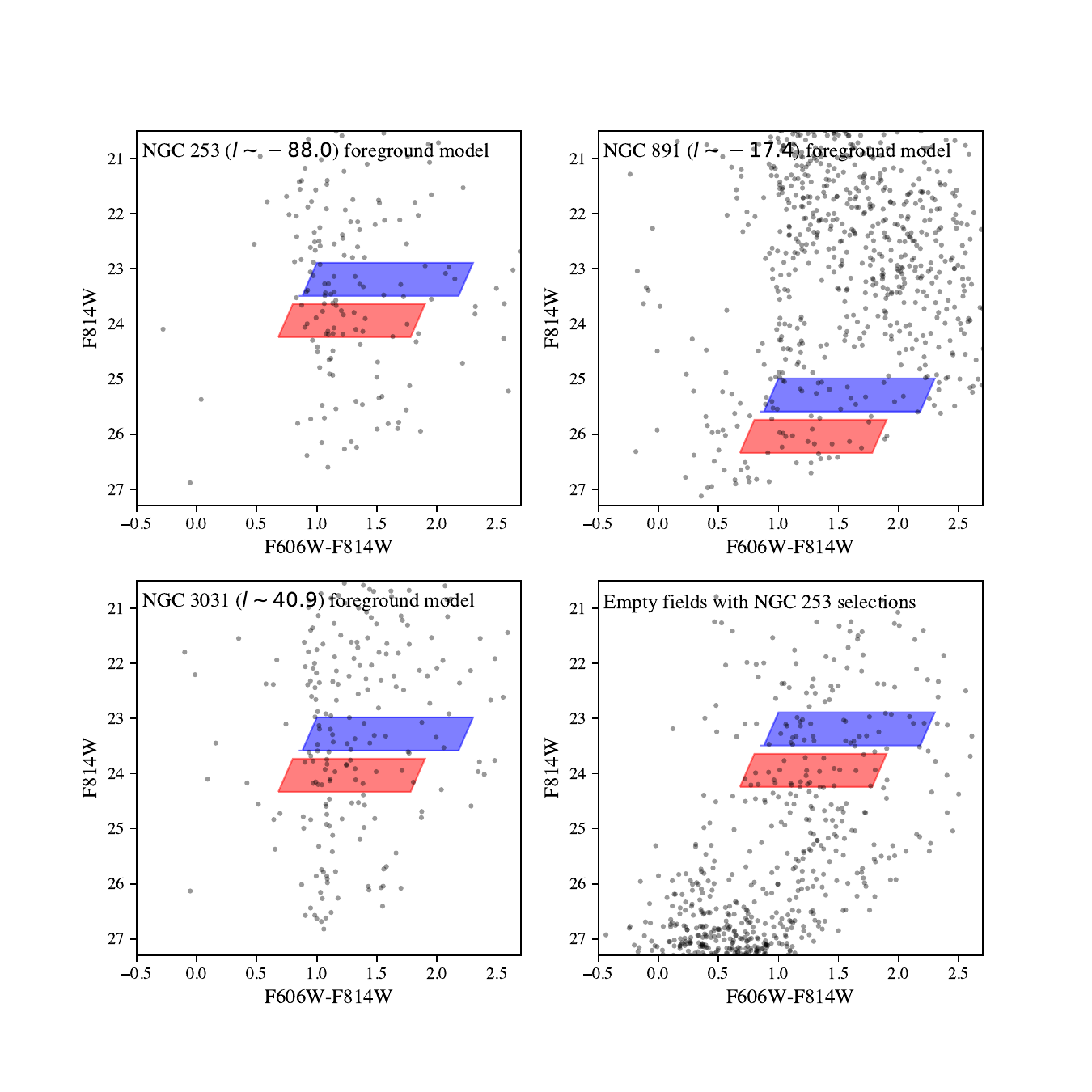}
\caption{Foreground model CMDs in the directions of NGC 253, NGC 891 and NGC 3031, along with the stacked CMD of our 8 `empty' HST fields that we use for background subtraction. The expected degree of contamination from foreground and background in our selection regions is small, and is accounted for in our measurements and error bars. }
\label{fg_bg}
\end{figure*}

While uncertainties from field-to-field variations in the number of foreground Milky Way stars and unresolved background galaxies are included in the error budget of our AGB/RGB measurements for NGC 253, NGC 891 and NGC 3031, we provide mock CMDs from the TRILEGAL galactic structure model \citep{Trilegal} in 0.025 square degrees regions (the equivalent of 8 ACS fields in size) in the directions of NGC 253, NGC 891 and NGC 3031, along with the combined CMD of our 8 empty fields in Fig.\ \ref{fg_bg}. The selection region on the CMD is overlaid. While the influence of the galactic latitude on foreground star counts is clear, the contamination of the selection regions is modest and does not affect our results. 

\section{Observational calibration of a relationship between AGB/RGB ratio and $t_{90}$} \label{ap:obscal}

\begin{table}\centering
    \begin{tabular}{lccl}
    
    Galaxy      &AGB/RGB       &$t_{90}$/Gyr & References \\
    \hline 
    Fornax & $0.157^{+0.021}_{-0.020}$ & $2.4^{+0.9}_{-0.3}$ & 1, 2 \\
    NGC 147 & $0.196^{+0.081}_{-0.073}$ & $2.7^{+6.4}_{-0.5}$ & 3, 2 \\
    NGC 185 & $0.155^{+0.065}_{-0.063}$ & $3.6^{+5.5}_{-0.5}$ & 3, 2 \\
    Cassiopeia III & $0.171^{+0.063}_{-0.050}$ & $4.1^{+2.5}_{-1.5}$ & 4, 5 \\
    Andromeda II & $0.097^{+0.049}_{-0.036}$ & $6.3^{+0.5}_{-0.6}$ & 6, 5 \\
    Andromeda I & $0.071^{+0.037}_{-0.030}$ & $7.4^{+0.9}_{-0.7}$ & 6, 5 \\
    Andromeda III & $<0.053$ & $8.7^{+1.5}_{-0.6}$ & 6, 5 \\
    Sculptor & $0.037^{+0.089}_{-0.037}$ & $10.7^{+1.3}_{-2.8}$ & 7, 2 \\
    Antlia &  $0.207^{+0.056}_{-0.044}$ & 2.32 & 8, 9 \\
    FM1/F6D1 & $ 0.090^{+ 0.018}_{-0.015}$ & 5.35 & 8, 9 \\
    Sc22/Sc-dE1 & $0.141^{+0.025}_{-0.022}$ & 3.42 & 8, 9 \\
    IKN & $ 0.082 ^{+0.009}_{-0.008}$ & 6.17 & 8, 9 \\ ESO 294-010 & $0.103^{+0.025}_{-0.020}$ & 3.78 & 8, 9 \\ 
    ESO 540-032 & $0.093^{+0.015}_{-0.013}$ & 3.45 & 8, 9 \\
    KDG 2 & $0.237^{+0.052}_{-0.043}$ & 0.68 & 8, 9 \\ 
    KK 77 & $0.130^{+0.011}_{-0.010}$ & 3.62 & 8, 9 \\
    ESO 410-005 & $0.147^{+0.023}_{-0.020}$ & 1.85 & 8, 9 \\
    HS 117 & $0.150^{ +0.024}_{-0.021}$ & 0.811 & 8, 9 \\
    KDG 63 & $0.139^{+0.012}_{-0.011}$ & 1.59 & 8, 9 \\ 
    UGC 8833 & $0.206^{+0.029}_{-0.025}$ & 1.20 & 8, 9 \\
    KDG 64 & $0.147^{+0.012}_{-0.011}$ & 1.41 & 8, 9 \\
    KDG 61 & $0.121^{+0.010}_{-0.009}$ & 1.30 & 8, 9 \\
    DDO 181 & $0.193^{+0.018}_{-0.016}$ & 1.12 & 8, 9 \\
    KKH 98 & $0.167^{+0.040}_{-0.032}$ & 1.23 & 8, 9 \\
    KDG 73 & $0.184^{+0.043}_{-0.035}$ & 1.74 & 8, 9 \\
    KKH 37 & $0.140^{+0.019}_{-0.017}$ & 1.54 & 8, 9 \\
    UGCA 292 & $0.152^{+0.051}_{-0.038}$ & 0.59 & 8, 9 \\
    DDO 113 & $0.122^{+0.020}_{-0.017}$ & 1.53 & 8, 9 \\
    DDO 44 & $0.147^{+0.014}_{-0.013}$ & 1.45 & 8, 9 \\
    GR8 & $0.155^{+0.032}_{-0.027}$ & 1.64 & 8, 9 \\
    DDO 78 & $0.121^{+0.010}_{-0.009}$ & 1.37 & 8, 9 \\
    DDO 6 & $0.150^{+0.021}_{-0.019}$ & 1.41 & 8, 9 \\
    UGC 8508 & $0.176^{+0.016}_{-0.015}$ & 1.11 & 8, 9 \\
    NGC 3741 & $0.170^{+0.018}_{-0.016}$ & 1.104 & 8, 9 \\ 
    NGC 4163 & $0.173^{+0.010}_{-0.009}$ & 6.13 & 8, 9 \\
    KDG 52 & $0.153^{+0.034}_{-0.028}$ & 2.04 & 8, 10 \\
    DDO 53 & $0.213^{+0.017}_{-0.016}$ & 1.19 & 8, 9 \\
    \hline 
    \end{tabular}
    \caption{AGB/RGB ratios and $t_{90}$ values for the observational sample of calibrating galaxies. References: 1) \citet{deBoer2012}, 2) \citet{Weisz2015}, 3) Hubble Source Catalog photometry WFC3 F606W/F814W (GO-15336; PI: A. Ferguson), 4) \citet{Martin2017}, 5) \citet{Weisz2019}, 6) \citet{Skillman2017}, 7) \citet{deBoer2011}, 8) \citet{Dalcanton2009}, 9) \citet{Weisz2011}, 10) \citet{Weisz2008}. }
    \label{Obs_calibration}
\end{table}

{Here, we present the galaxies that were used to calibrate the relationship between AGB/RGB ratio and $t_{90}$ in Table \ref{Obs_calibration}. As described in the text, we join two samples of galaxies towards this goal. Our first galaxies are a sample of relatively massive satellites in the Local Group that are close enough that accurate SF histories exist have been inferred from deep CMDs \citep{Weisz2015,Weisz2019}: Fornax, NGC 147, NGC 185, Cassiopeia III, Andromedas I, II and III, and Sculptor. Fornax and Sculptor have ground-based $V$- and $I$-band imaging \citep{deBoer2011,deBoer2012}: the translation between ground-based V and I-band photometry and F606W and F814W is taken from \citet{Rejkuba2005}. The cuts applied to select AGB and RGB stars are the same as for the GHOSTS halo observations and models. Because of the multiple observation types and data sources, background fields were not readily available. Recognizing that the AGB contamination is dominated by MW foreground halo stars, we very roughly estimate the AGB background from a brighter selection --- too bright to include AGB stars from the galaxy of interest --- covering the same colour range. The considerable uncertainties inherent to this method are incorporated in the error bars for the observational AGB/RGB ratios. }

{The Local Group has relatively few galaxies with the combination of sufficient luminosity to have a sizeable AGB star population and published SFH information. We therefore augmented the Local Group sample with a sample of Local Volume galaxies with published SFH information from \citet{Weisz2008} and \citet{Weisz2011} and published ACS photometry in any two of the F475W, F555W, F606W and F814W filters from \citet{Dalcanton09}. AGB/RGB ratios were measured assuming no background subtraction; error bars reflect counting statistics alone. Values of $t_{90}$ were interpolated from tabulated SFHs in \citet{Weisz2011} and have no reported error bars. For KDG 52, we choose to adopt the SFH from \citet{Weisz2008}, as KDG 52's CMD (especially the region bluewards of the RGB) appears to be much more similar to galaxies with more prominent intermediate age and young populations than the SFH from \citet{Weisz2011} would indicate, instead appearing more similar to galaxies with SFHs similar to that presented in \citet{Weisz2008}.  These galaxies tend towards much lower values of $t_{90}$ and higher AGB/RGB ratios than the Local Group sample, populating that part of the relation but not adding to the set of galaxies with older $t_{90} > 6$\,Gyr. In particular, many of the systems with $t_{90}<1.5$\,Gyr have some degree of ongoing star formation \citep{Weisz2011}, in agreement with model expectations of $t_{90}\sim1$\,Gyr or less for such systems.  This includes the most apparent outlier from this sample: NGC 4163 had most of its star formation $>6$\,Gyr ago, but has strong evidence of a significant episode of star formation $\sim 1$\,Gyr ago that produces a prominent AGB and an AGB/RGB ratio characteristic of systems with much lower $t_{90}$; indeed, $t_{90}<1$\,Gyr is within NGC 4163's 68\% confidence interval \citep{Weisz2011}. }

\label{lastpage}
\end{document}